\documentclass{aa}
\usepackage{natbib}
\usepackage{graphicx}
\usepackage{txfonts}
\usepackage{hyperref}
\begin{document}

   \title{A catalog of near-IR absolute magnitudes of Solar System small bodies}

   \author{Alvaro Alvarez-Candal
          \inst{1,2}
          \and
          Juan Luis Rizos\inst{1}
          \and
          Milagros Colazo\inst{3}
          \and
          Ren\'e Duffard\inst{1}
          \and
          David Morate\inst{4}
          \and
          Valerio Carruba\inst{5,6}
          \and
          Julio I.B. Camargo\inst{7,6}
          \and
          Andr\'es G\'omez-Toribio\inst{8}
          %\inst{4,5}%\fnmsep\thanks{Just to show the usage
          %of the elements in the author field}
          }

         \institute{Instituto de Astrof\'isica de Andaluc\'ia, CSIC, Apt 3004, E18080 Granada, Spain\\
              \email{alvaro@iaa.es}
        \and
        Instituto de F\'isica Aplicada a las Ciencias y las Tecnolog\'ias, Universidad de Alicante, San Vicent del Raspeig, E03080, Alicante, Spain
        \and
        Astronomical Observatory Institute, Faculty of Physics and Astronomy, A. Mickiewicz University, Słoneczna 36, 60-286 Poznań, Poland
        \and
        Centro de Estudios de F\'isica del Cosmos de Arag\'on
        \and
        UNESP, School of Engineering and Sciences, Department of Mathematics, Av. Dr. Ariberto Pereira
        da Cunha, 333, Guaratinguetá, 12516-410, São Paulo, Brazil
        \and
        Laboratório Interinstitucional de e-Astronomia - LIneA, Av. Pastor Martin Luther King Jr, 126 - Del
        Castilho, Rio de Janeiro, 20765-000, Brazil
        \and
        Observatório Nacional, Rua Gal. José Cristino 77, Rio de Janeiro, RJ - 20921-400, Brazil
        \and
        Escuela Técnica Superior de Ingeniería Informática (ETSINF), Universitat Politècnica de València, Valencia, Spain}

   \date{Received XX; accepted YY}

% \abstract{}{}{}{}{}
% 5 {} token are mandatory
 
  \abstract
{Phase curves of small bodies are useful tools to obtain their absolute magnitudes and phase coefficients. The former relates to the object's apparent brightness, while the latter relates to how the light interacts with the surface. Data from multi-wavelength photometric surveys, which usually serendipitously observe small bodies, are becoming the cornerstone of large statistical studies of the Solar System. Nevertheless, to our knowledge, all studies have been carried out in visible wavelengths.}
{We aim to provide the first catalog of absolute magnitudes in near-infrared filters (Y, J, H, and K). We will study the applicability of a non-linear model to these data and compare it with a simple linear model.}
{We compute the absolute magnitudes using two photometric models: the HG$_{12}^*$ and the linear model. We employ a combination of Bayesian inference and Monte Carlo sampling to calculate the probability distributions of the absolute magnitudes and their corresponding phase coefficients. We use the combination of four near-infrared photometric catalogs to create our input database.}
{We produced the first catalog of near-infrared magnitudes. We obtained absolute magnitudes for over 10\,000 objects (with at least one absolute magnitude measured), with about 180 objects having four absolute magnitudes.
We confirmed that a linear model that fits the phase curves produces accurate results. Since a linear behavior well describes the curves, fitting to a restricted phase angle range (in particular, larger than 9.5 deg) does not substantially affect the results.
Finally, we also detect a phase-coloring effect in the near-infrared, as observed in visible wavelengths for asteroids and trans-Neptunian objects.}
{}


   \maketitle

\section{Introduction} \label{sec:intro}

Phase curves of small bodies are valuable tools to compute absolute magnitudes and, through suitable photometric models, phase coefficients that may be related to surface properties. In a nutshell, the phase curve of a small body shows the change of apparent magnitude (normalized to unit distances object-Sun, $r$, and object-observer, $\Delta$) with phase angle ($\alpha$). The phase angle is related to the fraction of illuminated surface as seen by the observer.

Absolute magnitudes $(H)$ are related to the diameter and albedo of the small body \citep{bowelllumme1979}. They aim to quantify the amount of light scattered by the object in contrast to the amount of light it emits, which occurs in thermal infrared wavelengths. On the other hand, the phase coefficients depend on the chosen photometric model. There is a variety of them, but for the sake of simplicity, and because our input data is relatively sparse (see Sect. \ref{sec:data}), we will use models with few parameters (Sect. \ref{sec:processing}), namely the HG$_{12}^{*}$ \citep{penti2016HG} and a linear model, both having only two free parameters. The first is an adaptation of the HG$_1$G$_2$ \citep{muinonen2010HG1G2} that is better suited for working with sparse data, while the second is the more straightforward form a phase curve could adopt.

Phase curves have been studied mainly in the visible range (500 to 900 nm), first using traditional observing campaigns \citep[for instance,][]{belskayashev2000, shev2016} and more recently using large photometric surveys or massive databases \citep[for example,][among others]{oszki2011pcs,veres2015,colazo2021}. These works used photometric measurements in a single filter, often Johnson's V, thus missing the nuances of how the spectra, or photo-spectra, behave with changing phase angles. A {\em typical} phase curve shows a linear behavior for $\alpha>7 - 10$ deg. Some objects also show an increase in brightness at low $\alpha$, especially below five degrees, called the Opposition Effect (OE), which may be due to a decrease in the inter-particle shadows \citep{hapke1963JGR} or coherent back-scattering \citep{muino1989OE}.

The logical extension of single-wavelength studies is the use of multi-wavelength data, which allows us to mimic the reflectance spectra of small bodies. Among the first systematic studies are the works by \cite{mahlke2021,alcan2022}, and \cite{alcan2024AA}, which performed analyses of multi-wavelength phase curves, demonstrating their importance. For their most interesting results we can mention the apparent changes in the phase coefficients with wavelength \citep{mahlke2021,carry2024}, but that it should be interpreted with great care \citep{alcan2024PSS}, or the changes in spectral slope with $\alpha$, which show that some objects may suffer bluing {\it and} reddening in different phase angle intervals \citep{wilawer2024MNRAS,alcan2024AA,colazo2025Icarus}.

Unfortunately, to our knowledge, no systematic effort has been made to study the phase curves of small bodies in near-infrared, NIR, wavelengths, possibly because the curves are not expected to show a strong OE \citep{sanchez2012} and because NIR observations are more challenging than in the visible, particularly when moving to the redder part of the spectrum (H and K regions), where the brightness of the sky may outshine the small bodies, making their detection difficult.

The objectives of this work are: 1$^{\rm st}$, produce a database of absolute magnitudes in the near-infrared combining data from different photometric catalogs (see Sect. \ref{sec:data}); 2$^{\rm nd}$, better understand which photometric model will be more beneficial to analyze NIR phase curves; and 3$^{\rm rd}$, check the impact of having phase curves on limited $\alpha$ coverage, in particular when low phase angle data is not available. We will use data from four catalogs that provide near-infrared magnitudes, including the Y, J, H, and K filters. We organized this work such that in the next Section, we describe the four databases used and the transformations from one to the other; in Sect. In Section \ref{sec:processing}, we describe the data processing. The last two Sections are devoted to presenting and discussing the results.

\section{Data} \label{sec:data}
As mentioned above, we utilized four catalogs of near-infrared magnitudes that cover the YJHK region ($\approx 1000 - 2400$ nm). Each catalog was obtained at different facilities, so the photometric systems are not identical. Therefore, we used suitable transformations found in the literature to minimize their systematic differences. We aimed to transform all surveys to the MOVIS system because it is the only one that covers all four filters. In this work, we use the terms ``filter'' and ``magnitude'' interchangeably, as there should be no confusion for the reader.

\begin{figure}[ht!]
\centering
\includegraphics[width=\hsize]{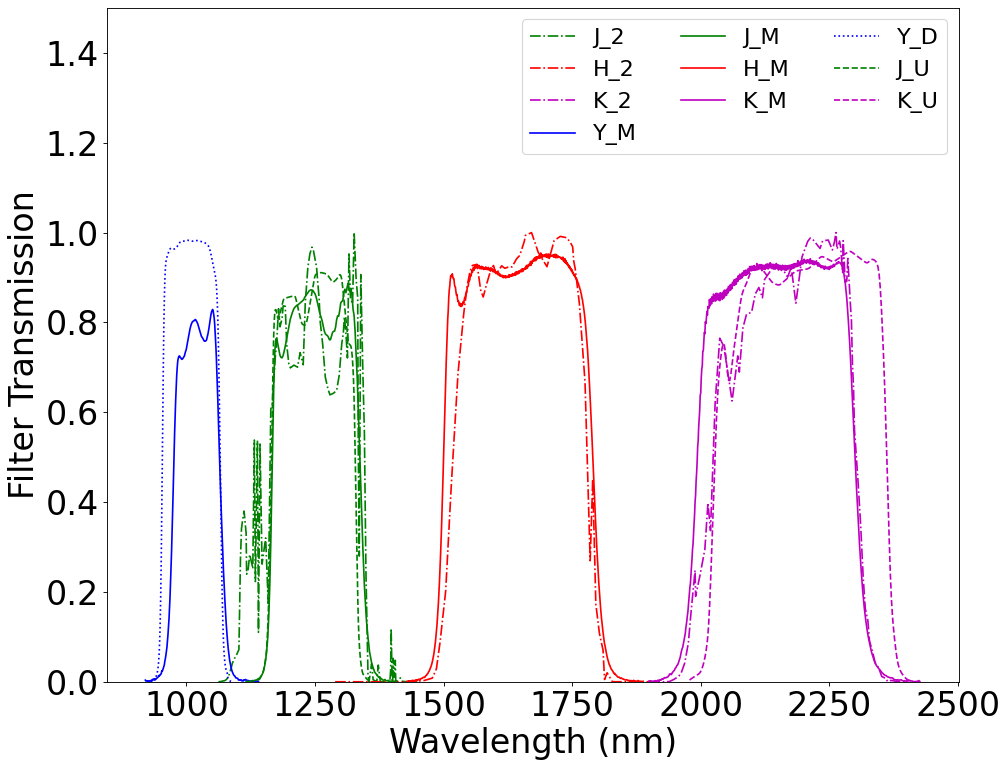}
\caption{Transmission curves of all filters in the near-infrared used in this work. Different colors label different filters, and different line styles indicate different instruments. The different labels are 2 for 2MASS, $M$ for MOVIS, $D$ for DES, and $U$ for UHS.}\label{fig:filters}%
\end{figure}

The following subsections describe the catalogs used and the corresponding transformations (also summarized in Appendix \ref{appB}). The total number of input objects can be seen in Table \ref{tab:objects}. The transmission curves of all filters can be seen in Fig. \ref{fig:filters}.
\begin{table}
\centering
\caption{Number of objects. Cells with no values are shown with (\ldots).}\label{tab:objects}
 \begin{tabular}{ lccccc}
 \hline
  & 2MASS & MOVIS & UHS & DES & All \\
 \hline
 \hline
Any & 19\,394  & 58\,502  & 13\,662  & 95\,802  & 169\,256 \\ 
Y   & (\ldots) & 42\,305  & (\ldots) & 95\,802  & 133\,710 \\ 
J   & 17\,594  & 57\,858  & 10\,323  & (\ldots) &  79\,403 \\ 
H   & 13\,993  & 18\,768  & (\ldots) & (\ldots) &  31\,539 \\ 
K   & 10\,138  & 34\,736  &  3\,689  & (\ldots) &  45\,474\\ 
 \hline
 \end{tabular}
\end{table}

\subsection{MOVIS}
The Moving Objects from VISTA survey \citep[MOVIS,][]{popescu2016movis} extracts the small body data from the VISTA Hemisphere Survey (VHS) carried out at the ESO-VISTA 4-m telescope in Cerro Paranal, Antofagasta, Chile. MOVIS collects data in the Y$_M$, J$_M$, H$_M$, and K$_M$ filters of 58\,502 objects (MOVIS-M catalog) with 73\,029 lines in total. The filter-by-filter breakdown is shown in Table \ref{tab:objects}. All details of matching the small bodies to the VISTA-VHS survey data can be found in \cite{popescu2016movis}. We will use this catalog as the reference because it includes data in all four filters. The data are available at the Centre de Donn\'ees de Strasbourg (\url{http://cdsarc.u-strasbg.fr/viz-bin/qcat?J/A+A/591/A115}).

\subsection{2MASS}
The 2 Microns All Sky Survey \citep[2MASS,][]{2mass2006AJ} observed almost the whole sky in three photometric filters J$_2$, H$_2$, and K$_2$, using two 1.3 m telescopes, one in Mount Hopkins, Arizona, USA; and one in Cerro Tololo, Coquimbo region, Chile. Among the almost $5\times10^8$ point-like sources cataloged, the survey detected several thousands of small bodies \citep[see][for details]{sykes20002mass}. The catalog is available in the Small Body Node of the Planetary Data System \citep{sykes2020}.

To transform the magnitudes from the 2MASS system to the MOVIS system, we used the equations in \cite{popescu2016movis}, which we repeat here (note that we use the sub-indexes $M$ and 2 for the MOVIS and 2MASS, respectively. We will keep this notation, which should be transparent to the reader, when necessary):
\begin{equation}\label{Eq:2mass2movis}
\begin{array}{l}
J_M = J_2 - 0.077 (J-H)_2,  \\
H_M = H_2 + 0.032 (J-H)_2,~{\rm and}  \\
K_M = K_2 + 0.010 (J-K)_2.  \\
\end{array}
\end{equation}

The transformations shown in Eq. \ref{Eq:2mass2movis} depend on the color of the object, which we do not know a priori. Moreover, not all objects were observed in all filters, nor simultaneously. Therefore, we decided to use solar colors in all cases (we used the same criterion in all transformations below). Suitable solar colors were chosen from the literature in the respective photometric system. In the case of 2MASS, we used the solar colors in \cite{casagrande2012ApJ}: $(J-H)_2 = 0.286$ and $(J-K)_2 = 0.362$. The offsets are $J_M-J_2\approx-0.02$, $H_M-H_2\approx0.01$, and $K_M-K_2\approx0.004$. The input 2MASS catalog contains 38\,280 lines\footnote{In this context, each line corresponds to a different observation.} for 19\,394 objects. The filter-by-filter breakdown is shown in Table \ref{tab:objects}.

\subsection{UHS}
The UKIRT Hemisphere Survey (UHS) currently observes the northern sky with three photometric filters: J$_U$, H$_U$, and K$_U$, although its current data release only includes J$_U$ and K$_U$ data. Within the survey, \cite{morrison2024AJ-UHS} extracted the data of small bodies serendipitously observed. The final catalog contains 15\,233 lines for 13\,662 objects.

The UHS magnitudes were transformed into MOVIS system using two steps. The first one, from \cite{hodgkin2009MNRAS}, transforms the magnitudes from UKIRT to 2MASS:
\begin{equation}\label{Eq:ukirt22mass}
\begin{array}{l}
J_2 = J_U + 0.065 (J-H)_U,  \\
H_2 = H_U + 0.030 - 0.070 (J-H)_U,~{\rm and}  \\
K_2 = K_U - 0.010 (J-K)_U,  \\
\end{array}
\end{equation}
\noindent
where we used the solar colors from \cite{willmer2018ApJS}: $(J-H)_U=0.35$ and $(J-K)_U = 0.40$. The offsets are $J_2-J_U\approx 0.02$, $H_2-H_U\approx0.05$, and $K_2-K_U\approx0.004$. Then, we applied the 2MASS $\rightarrow$ MOVIS transformation from Eq. \ref{Eq:2mass2movis}. The data are available as a machine-readable file in \url{https://iopscience.iop.org/article/10.3847/1538-3881/ad6909}.

\subsection{DES}
The Dark Energy Survey \citep[DES,][]{DES2016MNRAS} utilizes the DECAM camera and a set of filters spanning the visible spectrum (not used in this work) and the Y filter. DES observed 5\,000 square degrees of the southern sky using the 4-m Blanco Telescope in Cerro Tololo Observatory, Chile. Although its primary purpose is cosmological, thousands of small bodies were serendipitously observed \citep[e.g.,][]{berna2022DES,carruba2024MNRAS}. The Y filter catalog contains  135\,092 lines for 95\,802 objects\footnote{In this work, we do not use the data of the $\approx800$ trans-Neptunian objects detected by \cite{berna2022DES}.}. 

The DES survey is carried out using the AB photometric system, while the other used here are in the Vega system. This difference in reference systems introduces an offset in the zero points, which should be accounted for, especially when dealing with colors. We applied the transformation
\begin{equation}
\begin{array}{l}
Y_M = Y_D - 0.6,
\end{array}
\end{equation} 
from \cite{gonzalezmnras}. The DES data are available at \url{https://des.ncsa.illinois.edu/releases/other/asteroids}.

\smallskip
Joining all four catalogs, we end up with an input of 261\,634 lines for 169\,256 objects. 

\section{Data processing}\label{sec:processing}
The data were processed following the same structure as in our previous works \citep{alcan2022, alcan2024AA}, with adaptations to work with NIR data (for example, the transformation between the different photometric systems mentioned above). As in the cited works, the processing first creates or opens, if it exists, the probability distribution of rotational amplitudes, $P_A(\Delta m)$, for each small body $A$, and then uses it, convolved with the observational errors, to randomly extract 10\,000 points to fit the selected photometric model. The final products are the probability distributions $P_A(H,{\rm Coeffs})$ of each small body's absolute magnitude and phase coefficients. This processing primarily affects the uncertainty estimates, not the central values.

In this work, we will use two photometric models. First, the HG$_{12}^*$ model \citep{penti2016HG} and then a linear model where $M(\alpha)= H + \beta\alpha$. The HG$_{12}^*$ model is an adaptation of the IAU adopted HG$_1$G$_2$ model described by 
\begin{equation}\label{eq:pcurve}
M(\alpha)=H-2.5\log{[G_1\Phi_1(\alpha)+G_2\Phi_2(\alpha)+(1-G_1-G_2)\Phi_3(\alpha)]},
\end{equation}
where $M(\alpha)=m-5\log{(R\Delta)}$ is the reduced magnitude, $m$ is the apparent magnitude, $R(\Delta)$ is the helio(topo)centric distance, in AU, $H$ is the absolute magnitude, $\Phi_i$ are known functions of $\alpha$, and $G_i$ are the phase coefficients. In the HG$_{12}^*$ model $G_1$ and $G_2$ are parametrized by $G_{12}^*$ according to $G_1=0.84293649\times {G_{12}^*}$ and $G_2=0.53513350\times(1-{G_{12}^*})$. We will run three experiments: i) fit the HG$_{12}^*$ model, ii) fit the linear model, and iii) fit the linear model but including only $\alpha>9.5$ deg to accomplish the objectives mentioned in the Introduction.

The first exercise involves studying how well the HG$_{12}^*$ model, which is non-linear, fits the phase curves in the near-infrared, especially remembering that they may not show a strong non-linear behavior \citep[see][]{sanchez2012}, i.e., they may not display a strong OE. The second exercise will employ a more straightforward linear model and compare its results with those of the previous one. The last exercise aims to compare the absolute magnitudes obtained using the full range of $\alpha$ to those obtained with restricted coverage, in this case, to provide an idea of the impact of the lack of low-$\alpha$ data. This last case justifies the linear model because phase curves show a linear behavior for $\alpha\gtrsim7.5$ deg \citep{belskayashev2000}.

From the input catalog (comprising the four catalogs mentioned above), we select only objects with at least three different observations in any filter, with no other quality cuts, such that the error in the individual measurements is not larger than one magnitude. We also requested that the range covered by $\alpha$ be at least five degrees to have a minimum coverage of the phase curve and avoid spurious fits\footnote{Except for the linear $\alpha>9.5$ case, where we limited the minimum $\Delta\alpha$ to 3 deg to have a statistically significant number of objects.}. All results are available in electronic format. Table \ref{table:1} shows a sample of the results.
\begin{table*}
\caption{Sample of the catalog.}
\label{table:1}
\centering
\begin{tabular}{c| c c c | c |c c c | c c }
\hline\hline
(1) ID-MPC designation & $H_J$ & $H_J^-$ & $H_J^+$ & N & $G_J$ & $G_J^-$ & $G_J^+$ & $\alpha_{min}$ (deg) & $\Delta\alpha$ (deg) \\
\hline
 Ceres   - 00001& -8.0  &  -8.0   &  -8.0   & 1 &  -8.0   &-8.0   &-8.0  &   8.94 &  7.76 \\   
  Iris   - 00007& 4.089 &   4.193 &	  3.985 & 9 &  0.836  &0.290  &1.306 &  16.32 &  7.77 \\    
Psyche   - 00016& 4.291 &   4.433 &	  4.138 & 3 &  0.639  &0.190  &1.198 &   2.35 &  9.76 \\    
\hline
\hline
(2) ID-MPC designation & $H_{Jl}$ & $H_{Jl}^-$ & $H_{Jl}^+$ & N & $\beta_J$ (mag per deg)& $\beta_J^-$ & $\beta_J^+$ & $\alpha_{min}$ (deg) & $\Delta\alpha$ (deg) \\
\hline
 Ceres   - 00001&-8.0   &  -8.0   &  -8.0	& 1	&  -8.0    &-8.0   &-8.0 &   8.94 &  7.76 \\   
  Iris   - 00007& 4.424 &	4.966 &	  3.887 & 9	&   0.034  &0.010  &0.059&  16.32 &  7.77 \\    
Psyche   - 00016& 4.467 &	4.680 &	  4.256 & 3	&   0.041  &0.012  &0.071&   2.35 &  9.76 \\    
\hline
\hline
(3) ID-MPC designation & $H_{Je}$ & $H_{Je}^-$ & $H_{Je}^+$ & N & $\beta_{Je}$ (mag per deg)& $\beta_{Je}^-$ & $\beta_{Je}^+$ & $\alpha_{min}$ (deg) & $\Delta\alpha$ (deg) \\
\hline
 Ceres   - 00001&-8.0   & -8.0  & -8.0  & 1 &-8.0   & -8.0   & -8.0  & 8.94  &7.76\\
  Iris   - 00007& 4.419 & 4.971 & 3.887 & 9 & 0.034 &  0.010 &  0.058&16.32  &7.77\\
Psyche   - 00016&-8.0   &-8.0   &-8.0   & 3 &-8.0   & -8.0   & -8.0  & 2.35  &9.76\\
\hline
\end{tabular}
\tablefoot{The first column shows the ID of the object and its Minor Planets Center ASCII designation. $H_J$ is the median of the probability distribution, while $^-$ and $^+$ designate the absolute magnitude at the 16th and 84th percentiles. The number of observations considered is shown in the fifth column. $G_J$, $G_J^-$, and $G_J^+$ (or $\beta$) from column six to eight. The last two columns show the minimum $\alpha$ and its total span. Flag -8 indicates objects with less than three observations in a given filter or $\Delta\alpha<5$ deg, while flag -9 indicates that not enough valid solutions were obtained {(that is, less than 100)}. The complete catalog is available at the CDS\footnote{\url{https://cdsarc.u-strasbg.fr/}}, the Open Science Framework\footnote{\url{https://osf.io/cfxq2/}}, or upon request. (1) Indicates results obtained using the HG$_{12}^*$ model, (2) obtained with the linear model, and (3) with the linear model restricted to $\alpha_{min}>9.5$ deg. We simplified the notation of the $G_{12\lambda}^*$ for clarity.}
\end{table*}

\section{Results and discussion}
We present the results in the sections below. The first section describes the results using the HG$_{12}^*$ model, the second uses a linear model across the entire $\alpha$ range, and the last Section describes the results of a linear approach, considering $\alpha_{min}>9.5$ deg. 

\subsection{HG$_{12}^*$ model}
We determined $H$ in the four near-infrared filters using the model described in \cite{penti2016HG}. See Table \ref{tab:objectsHG} for the total number of magnitudes determined. In the Table, columns 2SMASS to DES show results where data from only one catalog were used. From column 2M onward, the results were obtained by combining data from different catalogs.
\begin{table*}
\centering
\caption{Number of objects with valid results. }\label{tab:objectsHG}
 \begin{tabular}{lrc|ccccccccc}
 \hline
Model            &    & Total   & 2MASS (2)& MOVIS (M)& UHS (U)  & DES (D)  & 2M       & 2U        & MU       & MD       & 2MU      \\
 \hline
 \hline
                 &Y   & 7\,730  & (\ldots) &  736     & (\ldots) & 5\,327   & (\ldots) & (\ldots)  & (\ldots) & 1667     & (\ldots) \\ 
HG$_{12}^*$      &J   & 4\,657  & 1\,207   &  1\,149  & 18       & (\ldots) & 1\,473   & 235       & 386      & (\ldots) &  189     \\ 
                 &H   & 1\,770  & 1\,114   &  201     & (\ldots) & (\ldots) & 455      & (\ldots)  & (\ldots) & (\ldots) & (\ldots) \\ 
                 &K   & 2\,150  &    508   &     627  & 16       & (\ldots) &    674   & 143       & 143      & (\ldots) &   72     \\ 
 \hline
                 &Y   & 7\,874  & (\ldots) &  771     & (\ldots) & 5\,421   & (\ldots) & (\ldots)  & (\ldots) & 1682     & (\ldots) \\ 
Linear           &J   & 5\,184  & 1\,450   &  1\,166  & 18       & (\ldots) & 1\,681   & 269       & 390      & (\ldots) &  210     \\ 
                 &H   & 2\,123  & 1\,336   &  219     & (\ldots) & (\ldots) & 538      & (\ldots)  & (\ldots) & (\ldots) & (\ldots) \\ 
                 &K   & 2\,519  &    668   &     642  & 16       & (\ldots) &    841   & 128       & 144      & (\ldots) &   80     \\ 
\hline
                 &Y   & 3\,173  & (\ldots) &  600     & (\ldots) & 1\,920   & (\ldots) & (\ldots)  & (\ldots) &  653     & (\ldots) \\ 
Linear           &J   & 2\,125  &    243   &  853     &  5       & (\ldots) &    617   &  54       & 134      & (\ldots) &   39     \\ 
$\alpha>9.5$ deg &H   &    761  &    344   &  217     & (\ldots) & (\ldots) &    200   & (\ldots)  & (\ldots) & (\ldots) & (\ldots) \\ 
                 &K   &    932  &    170   &  399     &  5       & (\ldots) &    272   &  26       &  46      & (\ldots) &   14     \\ 
\hline
 \end{tabular}
 \tablefoot{The first column shows the photometric model used, the second column the filter, and the third the total number of valid results. Column 2MASS shows valid results obtained using {\it only} 2MASS data, likewise for columns five, six, and seven, for MOVIS, UHS, and DES, respectively. Column eight shows results obtained combining data from 2MASS and MOVIS, likewise for columns nine to twelve for 2MASS and UHS, MOVIS and UHS, MOVIS and DES, and 2MASS, MOVIS, and UHS, respectively. Cells with no values are shown with (\ldots).}
\end{table*}
We obtained results for 11\,841 objects with at least one valid absolute magnitude (in any filter), while 143 have all four valid magnitudes.

\begin{figure}[ht!]
\centering
\includegraphics[width=\hsize]{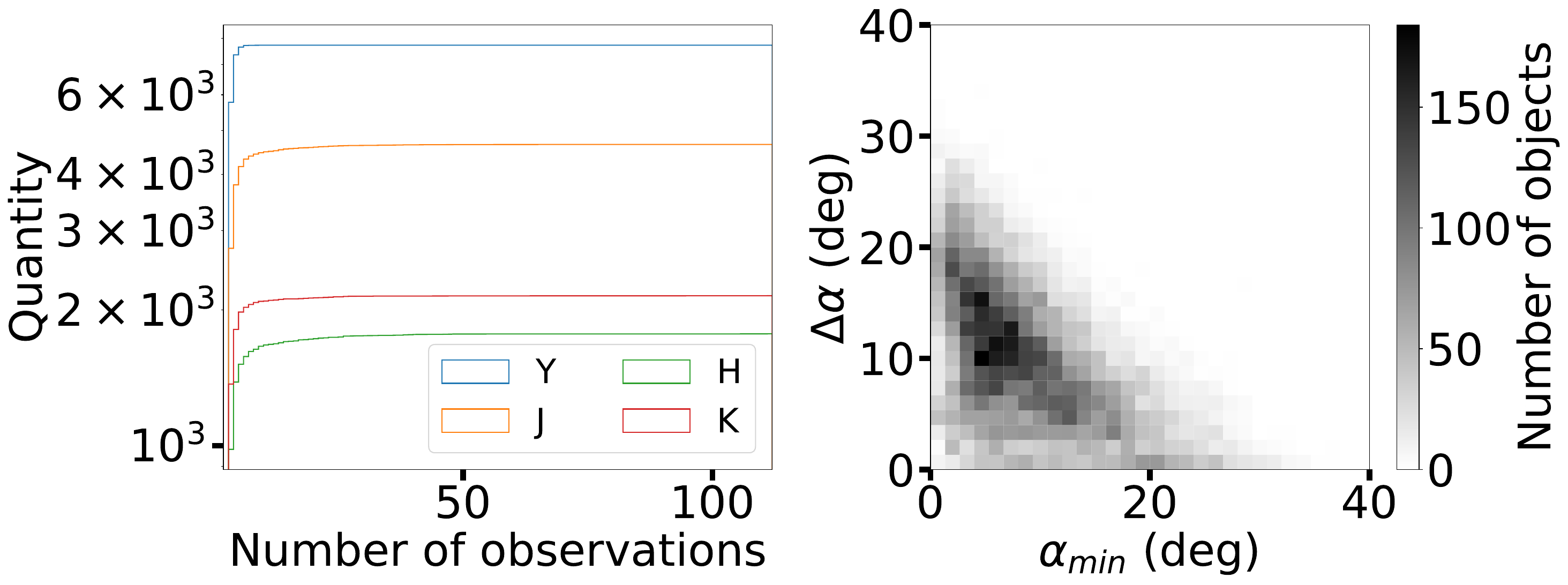}
\caption{Left panel: Cumulative distribution of absolute magnitudes determined filter by filter, shown in different colors. Right panel: Heat map showing the distribution of minimum $\alpha$ and total range of phase angle covered by the objects with measured absolute magnitudes in at least one filter.}\label{fig:obs}%
\end{figure}

Most results are in the Y filter, while the least numerous are in the H filter (see the right panel in Fig. \ref{fig:obs}).
Most results cover $\alpha_{min}\in(1,15)$ deg and $\Delta \alpha \in(10,20)$ deg (left panel in Fig. \ref{fig:obs}).

\begin{figure}
\centering
\includegraphics[width=4.4cm]{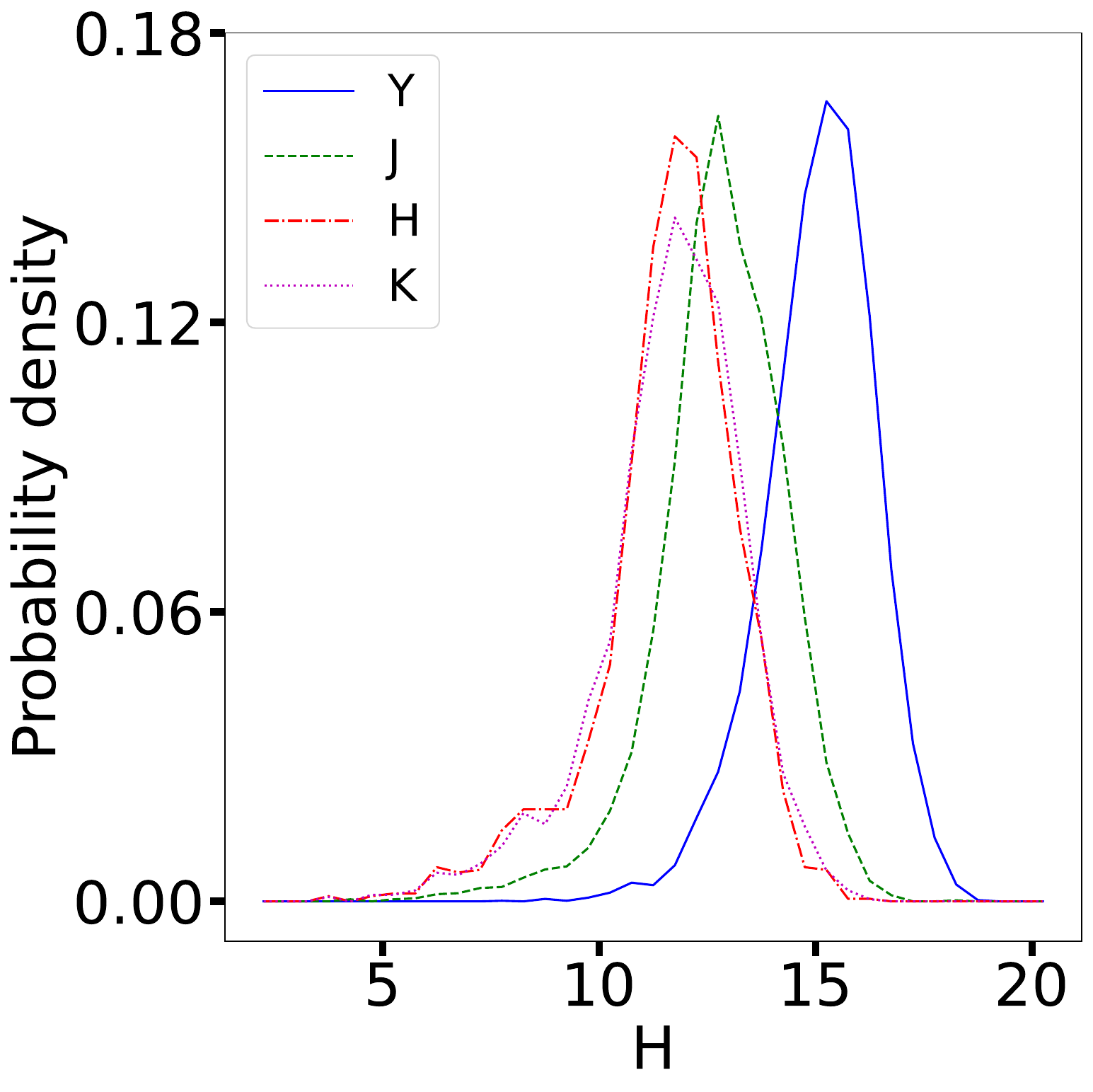}
\includegraphics[width=4.4cm]{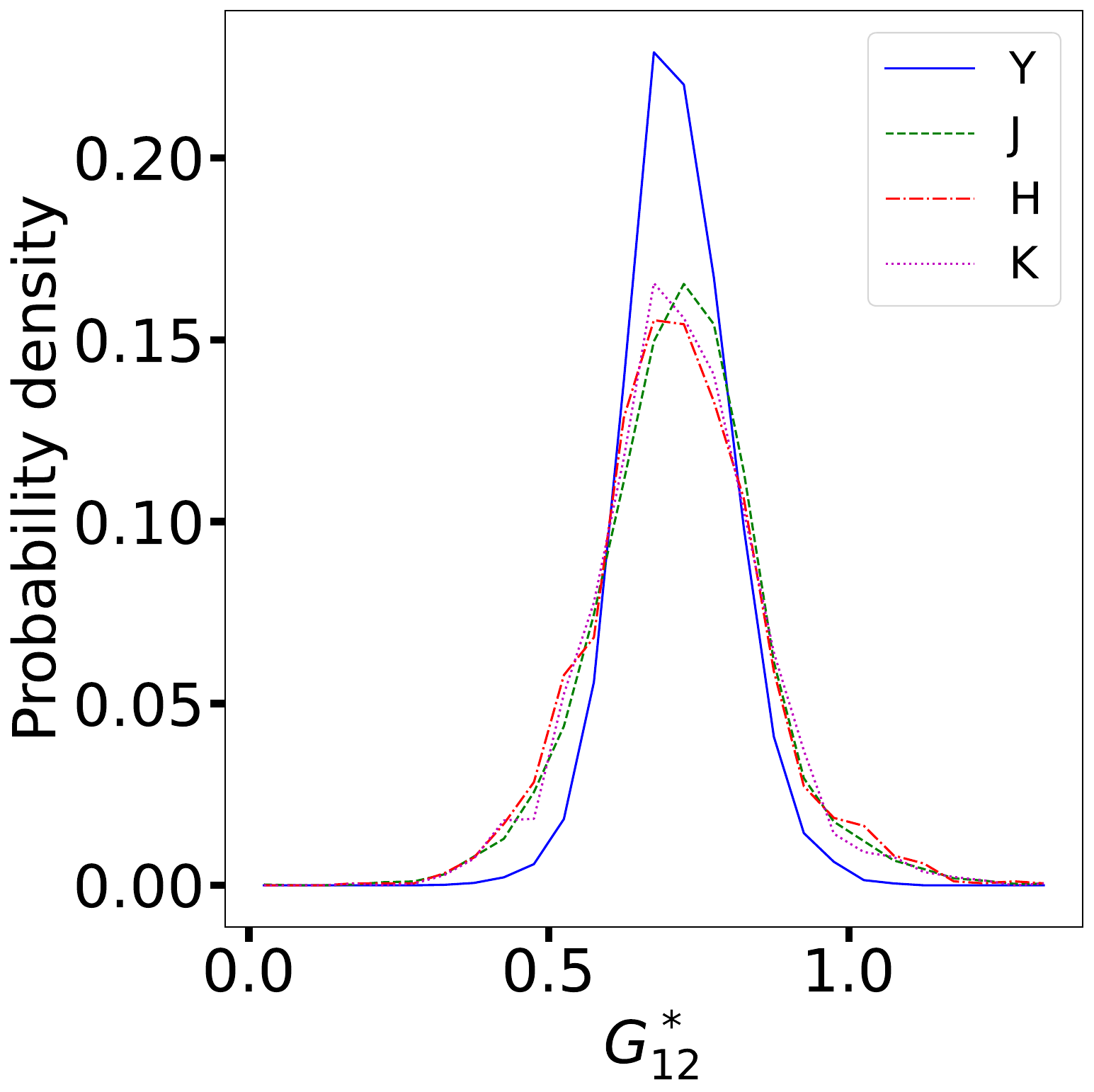}
\caption{Distributions of computed quantities. Distributions of $H$ (left panel) and $G_{12}^*$ (right panel). $Y$ is shown as the blue line, $J$ is shown in green, $H$ is shown in red, and $K$ is shown in purple.}\label{fig:disthg12}%
\end{figure}

Figure \ref{fig:disthg12} shows the distributions of absolute magnitudes and phase coefficients. The Y magnitude distribution peaks at fainter magnitudes than all the rest, with a median magnitude of 15.2. The median J magnitude is about one magnitude fainter than the median H and K magnitudes (12.9, 11.8, and 11.8, respectively). The H and K distributions are remarkably similar, showing some ripples at the bright end, between 5 and 10 magnitudes (Fig. \ref{fig:disthg12}, left panel). The phase coefficients all have median values between 0.70 and 0.72 and show similar behavior, except for the $G_{12Y}^*$, which is slightly sharper than the rest and, consequently, slightly higher. To facilitate comparison, all distributions are normalized to a unit area.
The visual similarity between the $G_{12}^*$ values may indicate that the phase coefficients are not strongly wavelength-dependent, at least population-wise. Nevertheless, to quantitatively assess this possible non-relation, we run the Kruskal-Wallis test, whose null hypothesis is that all distributions have the same median value. The null hypothesis can be rejected if $p_{value}<0.05$. The test shows a $p_{value}=8\times10^{-3}$, rejecting the null hypothesis. The Kruskal-Wallis test does not indicate which distributions are responsible for this. Therefore, we used the Post hoc Dunn test, as implemented in {\tt scikit\_posthocs}. The test revealed that the distribution driving the rejection is $G_{12Y}^*$. Still, it is unclear whether an actual wavelength dependence exists in the phase coefficients. As mentioned above, $H_Y$ reaches fainter magnitudes than the other distributions, which may be because asteroids tend to be brighter in the Y filter. Also, the sky background increases with wavelength, making it more difficult to observe. 

\begin{figure}
\centering
\includegraphics[width=4.4cm]{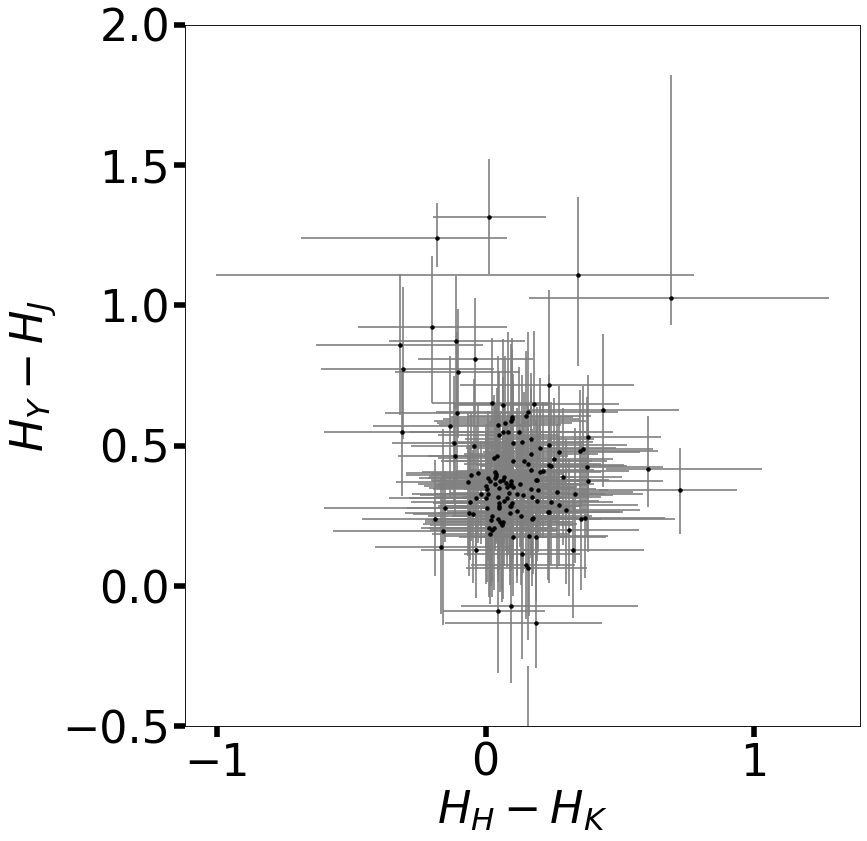}
\includegraphics[width=4.4cm]{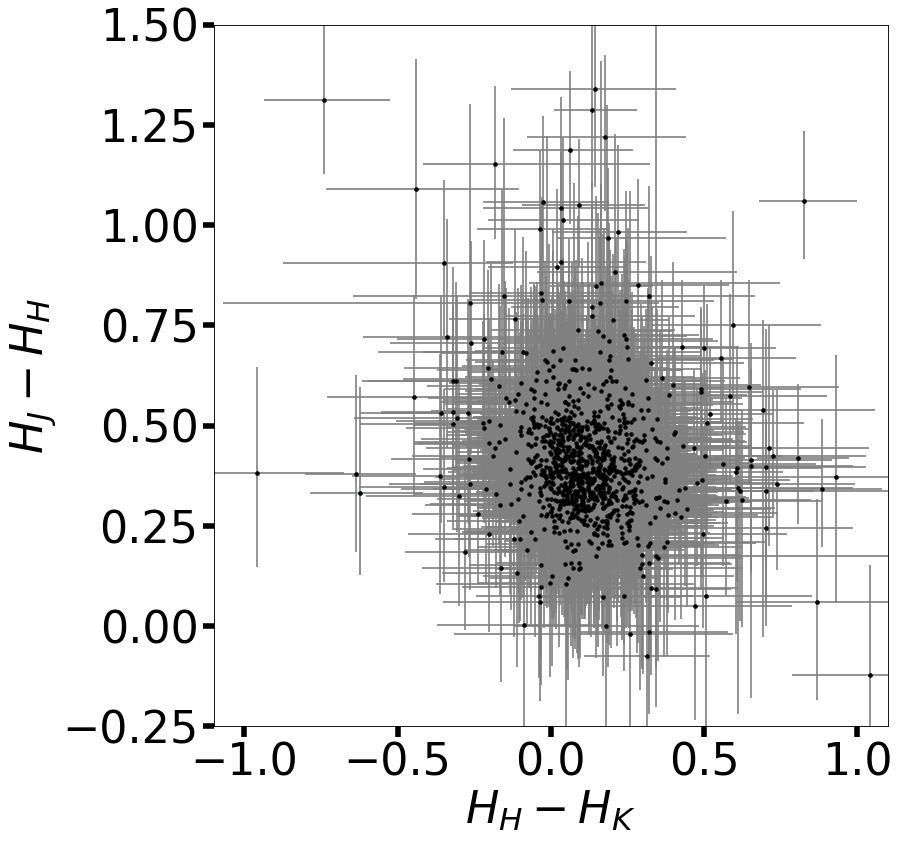}
\caption{Small bodies color-color plots. The left panel shows the $H_H - H_K$ vs $H_Y - H_J$, while the right panel shows the $H_H - H_K$ vs $H_J - H_H$.}\label{fig:colorshg12}%
\end{figure}

The color distribution of the absolute magnitudes, dubbed as absolute colors, constructed just by making the difference in the nominal values, can be seen in Fig. \ref{fig:colorshg12}.
The first noticeable detail is the remarkable difference in numbers between the two plots. All plots involving the $H_Y$ have fewer matches than the other filters. For example, the left panel includes 140 objects (plus one outlier outside the plot's limits), while the right panel includes 1035 (plus one outlier). There are no obvious groupings in any of the plots. The errors in the figure (and in the rest of the text) are estimated as half the distance between the 16th and the 84th percentiles of the distributions of the physical quantities.

In the next section, we present the results of the same input database that is processed through a linear photometric model.

\subsection{Impact of Photometric Model Choice}\label{sec:largea}
After inspecting the solutions obtained with the HG$_{12}^*$ model, we decided to test the simpler linear model of the form
\begin{equation}\label{eq:linear}
    M_{\lambda} = H_{l\lambda} + \beta_{\lambda}\alpha,
\end{equation}
where $H_{l\lambda}$ is the absolute magnitude and $\beta_{\lambda}$ is the phase coefficient in units of deg per mag. It is not only the simplicity of the model that is attractive (and its possibly larger number of valid solutions and quicker computation), but also the fact that we do not expect a strong OE in the NIR. Therefore, we ran the same input database with the linear model, obtaining results in at least one filter for 12\,394 (more than 500 more objects than using the HG$_{12}^*$ model) and with all four valid absolute magnitudes for 180 objects, 37 more than above. The increase does not seem large, but it implies a 5\% increase in the first number and a 20\% increase for all valid absolute magnitudes.

\begin{figure}
\centering
\includegraphics[width=4.4cm]{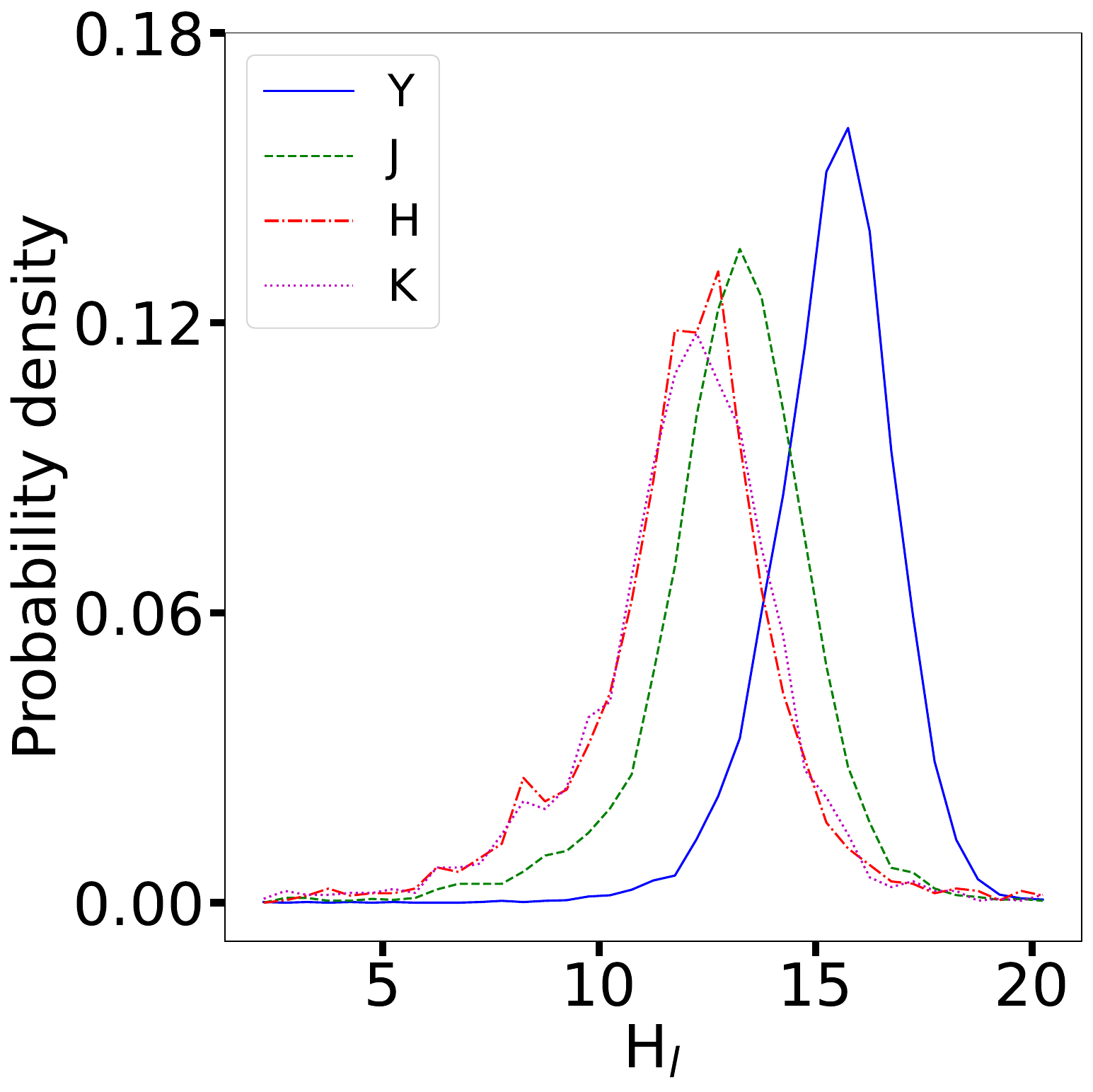}
\includegraphics[width=4.4cm]{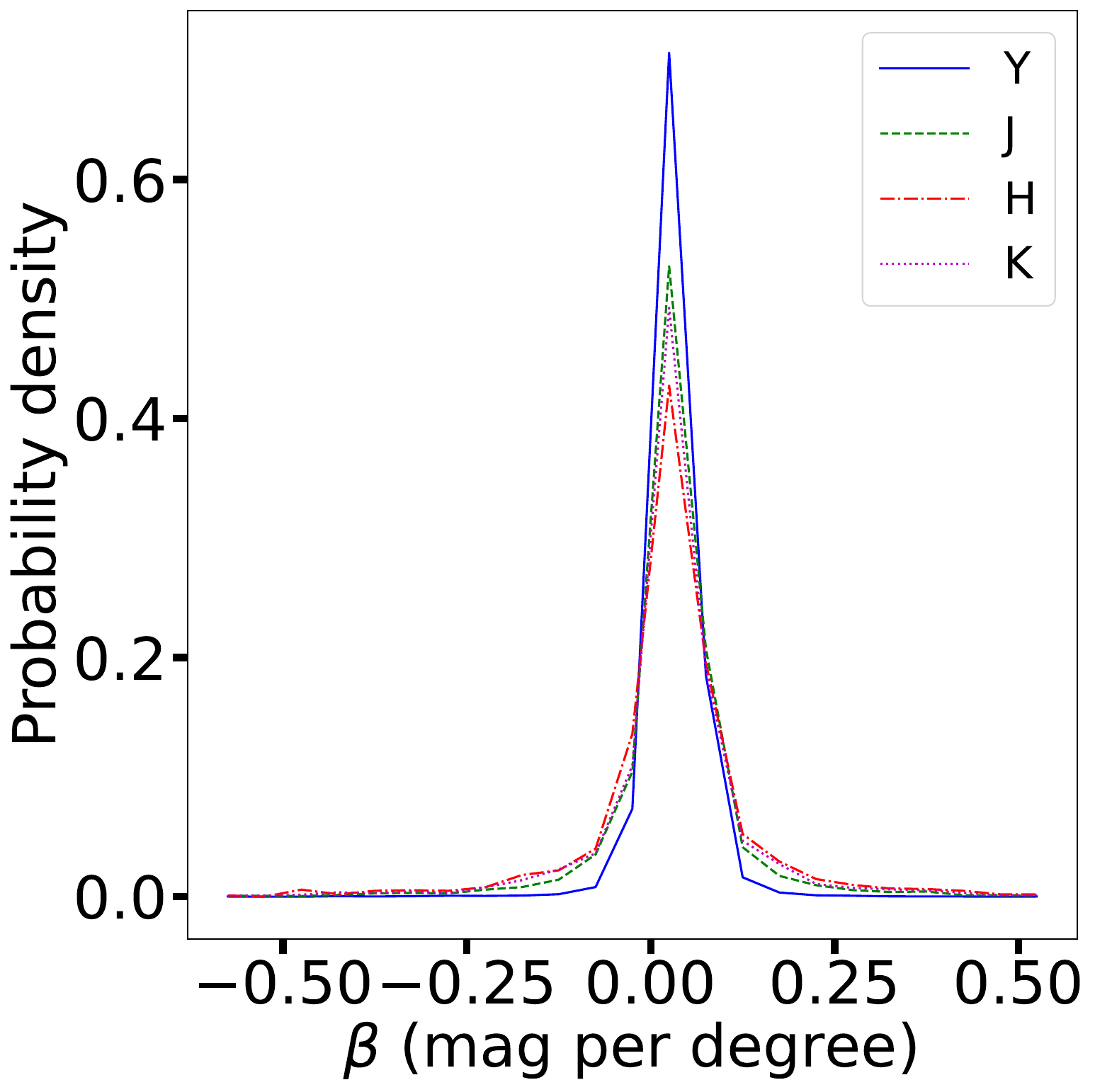}
\includegraphics[width=4.4cm]{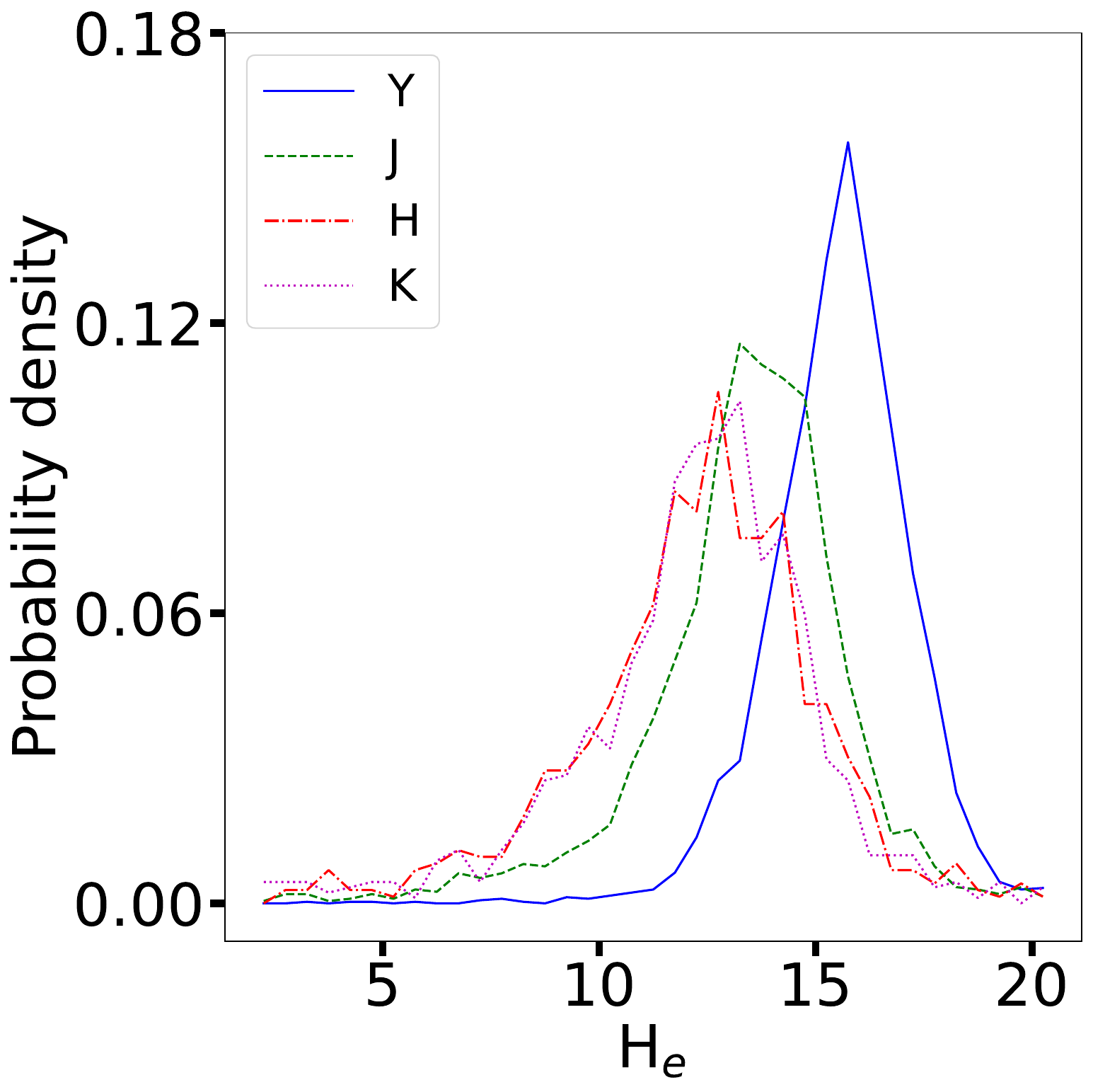}
\includegraphics[width=4.4cm]{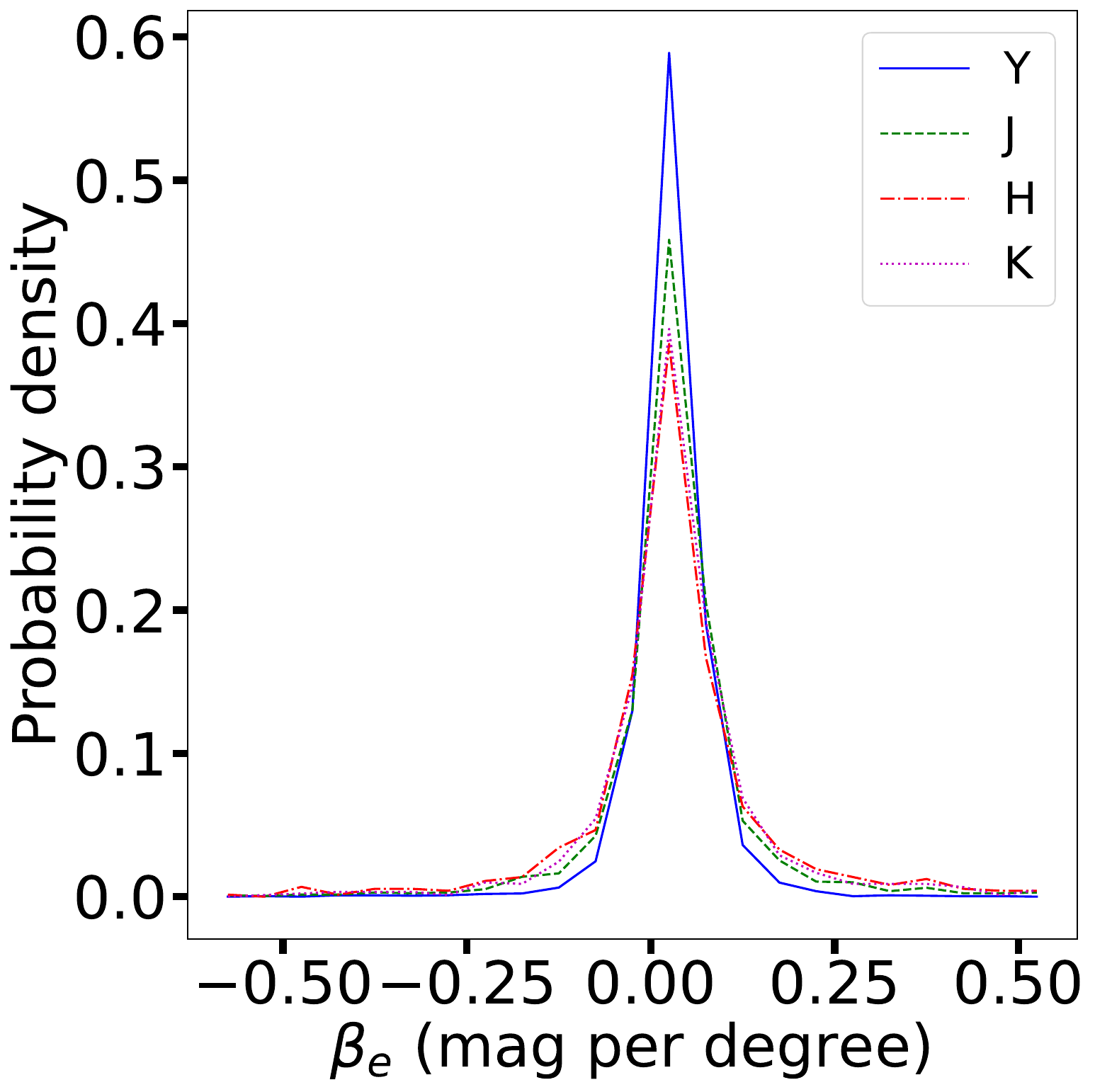}
\caption{Distributions of absolute magnitudes and linear phase coefficients. Top panels were obtained using the full range of $\alpha$. Bottom panels were obtained using $\alpha>9.5$ deg. The left column shows the absolute magnitude distributions. The right column shows the distributions of $\beta$. $Y$ is shown as the blue line, $J$ is shown in green, $H$ is shown in red, and $K$ is shown in purple.}\label{fig:distbetas}%
\end{figure}

Figure \ref{fig:distbetas}, top right panel, shows the $\beta_{\lambda}$ distribution. Despite the apparent similarities of all $\beta$ distributions, the Kruskal-Wallis test gives $p_{value}=0.024$, rejecting the null hypothesis that all median values are the same. The Dunn test indicates that the $\beta_Y$ and $\beta_J$ distributions differ with respect to the $\beta_H$ and $\beta_K$ distributions. All distributions have median values of 0.03 mag per deg, relatively thin and accepting a few negative values, with barely any strong outliers (11 in Y, 32 in J, 51 in H, and 30 in K with $|\beta_{\lambda}|>1.5$ mag per deg). The $H_{\lambda}$ distributions peak at 15.51 (Y), 13.22 (J), 12.23 (H), and 12.20 (K) (top left panel).
The negative values of $\beta$ may not be related to the scattering properties of the surfaces, and their explanation probably comes from macroscopic reasons such as an unfortunate sampling of the rotational light-curve, or faint cometary-like activity \citep{alcan2016,ayala2018}.

\begin{figure}[ht!]
\centering
\includegraphics[width=4.4cm]{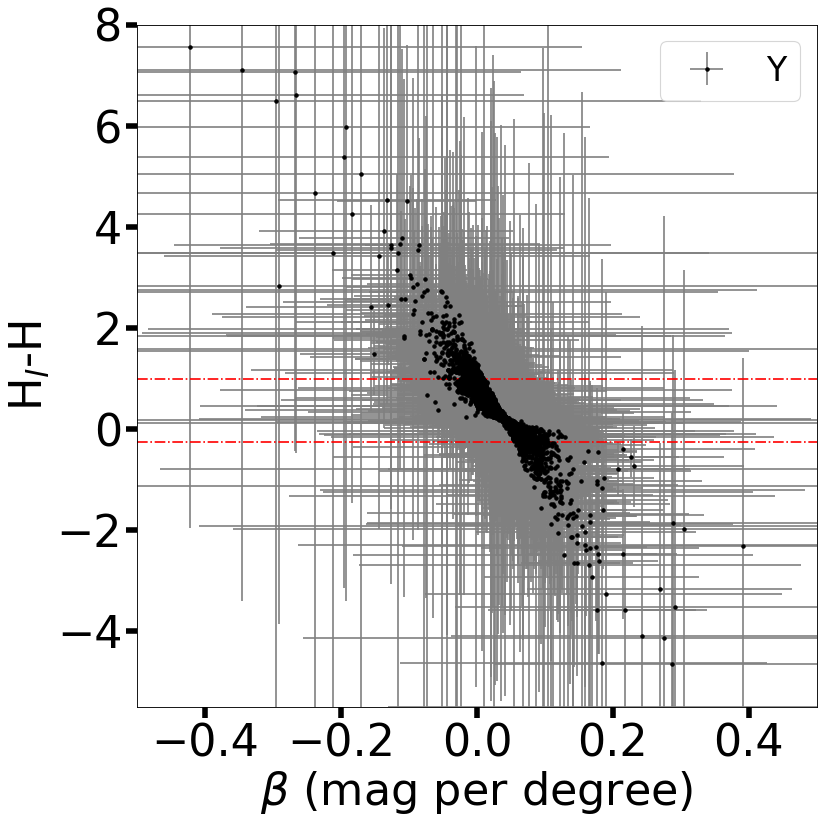}
\includegraphics[width=4.4cm]{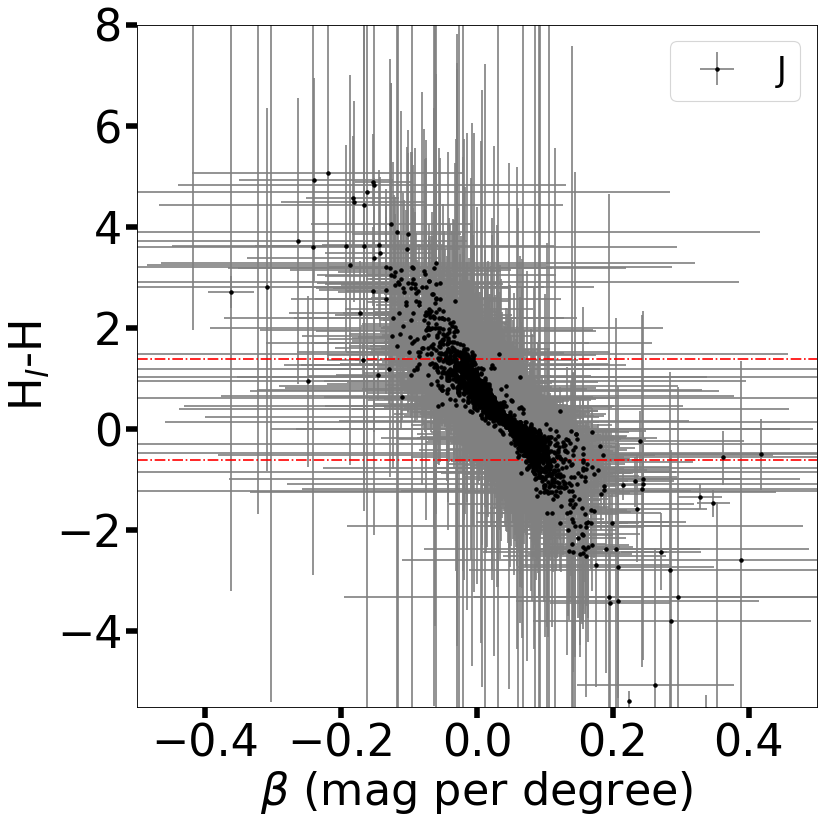}
\includegraphics[width=4.4cm]{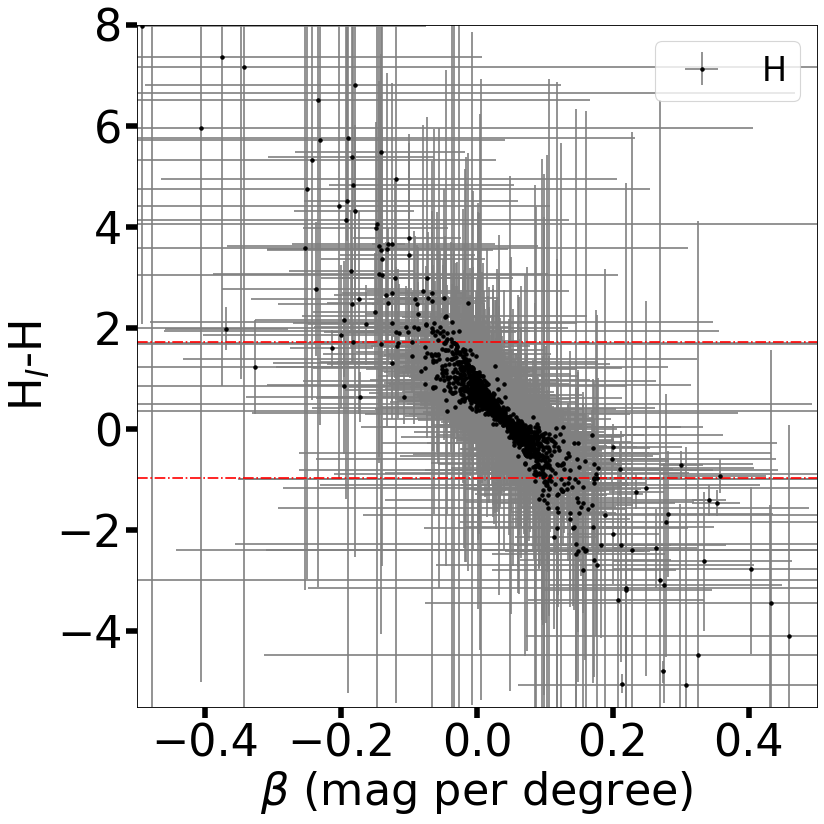}
\includegraphics[width=4.4cm]{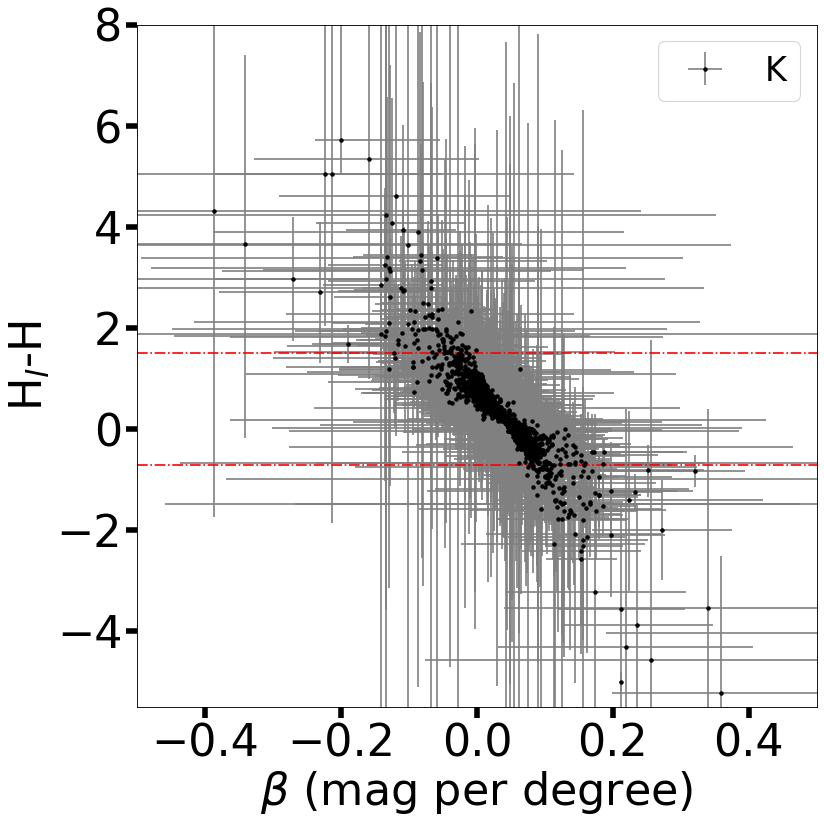}
\caption{Scatter plots showing the difference in the absolute magnitudes computed using the HG$_{12}^*$ model (H) and the linear model (H$_l$) versus the linear phase coefficient ($\beta$). The red dashed lines indicate where 90\% of the results lie.}
\label{fig:g12-beta}
\end{figure}

We evaluated the impact of using the different photometric models on the estimated absolute magnitudes, creating $H_{l}-H$ plots (as shown in Fig. \ref{fig:g12-beta}). The difference in the absolute magnitudes per filter is shown against the value of the corresponding $\beta$. All figures follow the same pattern, strongly influenced by the linear model's correlation between $\beta$ and $H_{l}$. The median difference in the Y and J filters is about 0.28, while in the H and K filters is about 0.3 mag. The horizontal lines show the range in magnitudes span by 90 \% of the sample: $\Delta H_{Y}\in[-0.27, 1.00]$, $\Delta H_{J}\in[-0.62, 1.38]$, $\Delta H_{H}\in[-0.94,1.70]$, and $\Delta H_{K}\in[-0.71,1.49]$. In all cases, the corresponding $\beta_{\lambda}$ values fall within -0.15 and 0.20 mag per degree (ignoring a few outliers).
These numbers may be interpreted in two different ways: (i) although the linear approximation is a good proxy to estimate the absolute magnitude in NIR filters, some effects are still not included in this simplistic model, or (ii) the linear model produces systematically fainter absolute magnitudes (Fig. \ref{fig:g12-beta}). Note that the difference in the absolute magnitudes is a measure of the OE, which is usually measured as the difference between the non-linear phase curve and a linear fit \citep[e.g.,][]{slyusarev2025}. Is it possible that the HG$_{12}^*$ model is overfitting the curves and producing an unreal OE?

Aiming for an answer, we directly compare the resulting models with the input magnitudes. Therefore, we performed two tests:
\begin{figure}[ht]
\centering
\includegraphics[width=4.4cm]{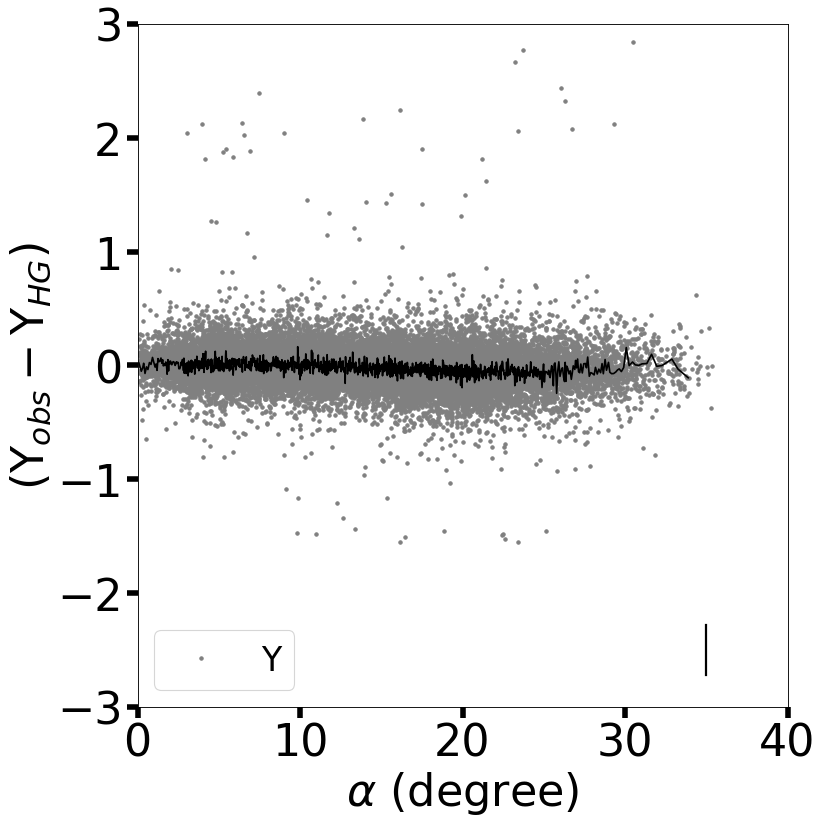}
\includegraphics[width=4.4cm]{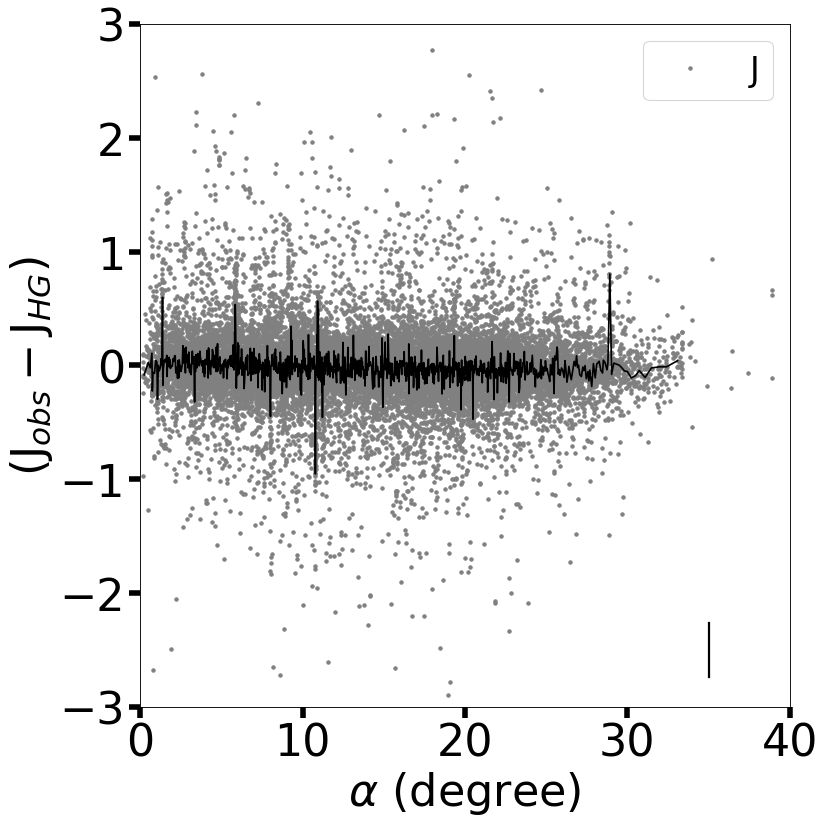}
\includegraphics[width=4.4cm]{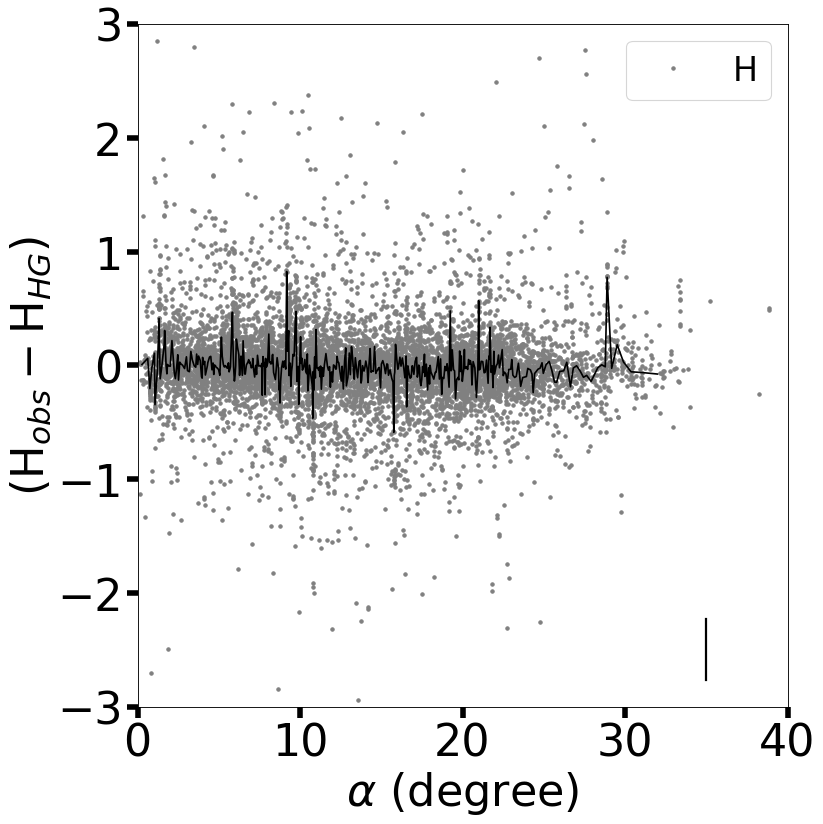}
\includegraphics[width=4.4cm]{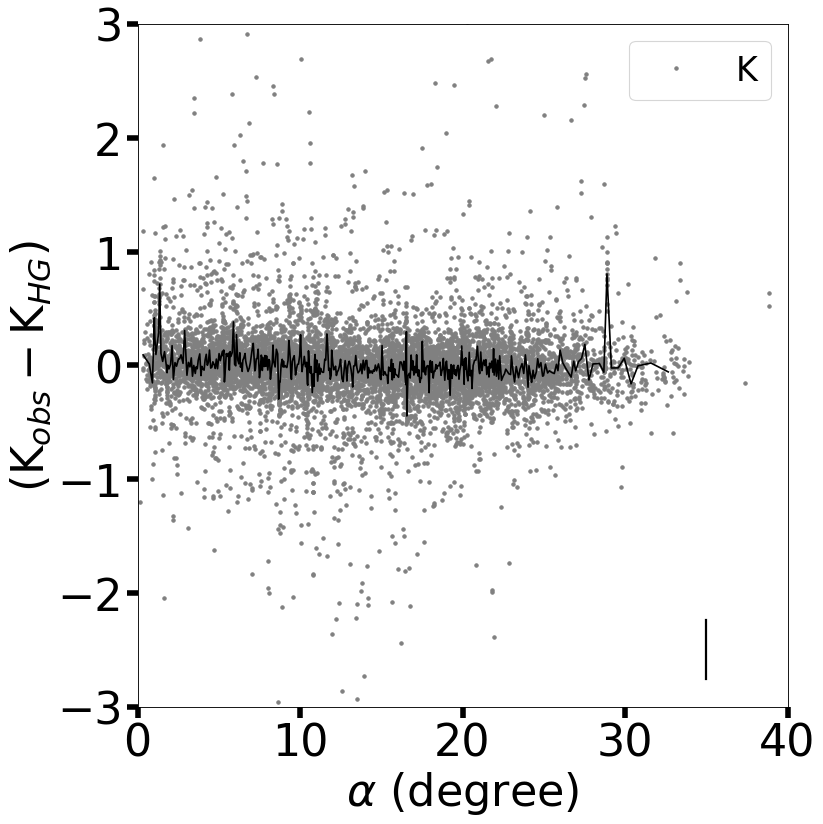}
\caption{Comparison between measured and modeled magnitudes using the HG$_{12}^*$ model. Different panels show data in different filters. The continuous line indicates a running median of 21 points. Typical error bars in the y-axis are shown in the lower-right corner for reference.}
\label{fig:comp1}
\end{figure}
\begin{figure}[ht]
\centering
\includegraphics[width=4.4cm]{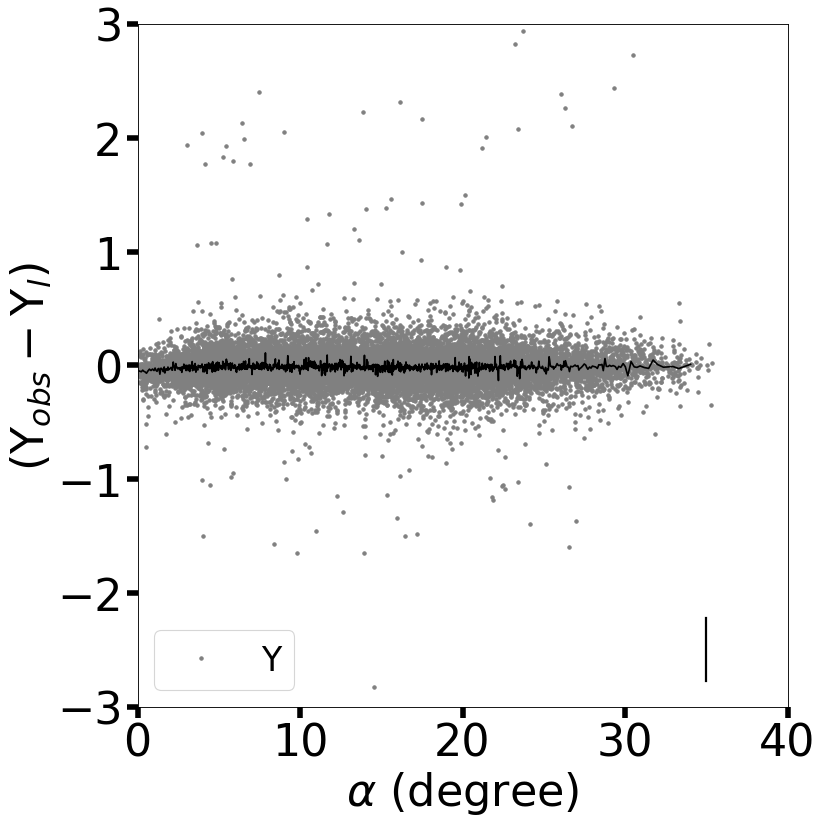}
\includegraphics[width=4.4cm]{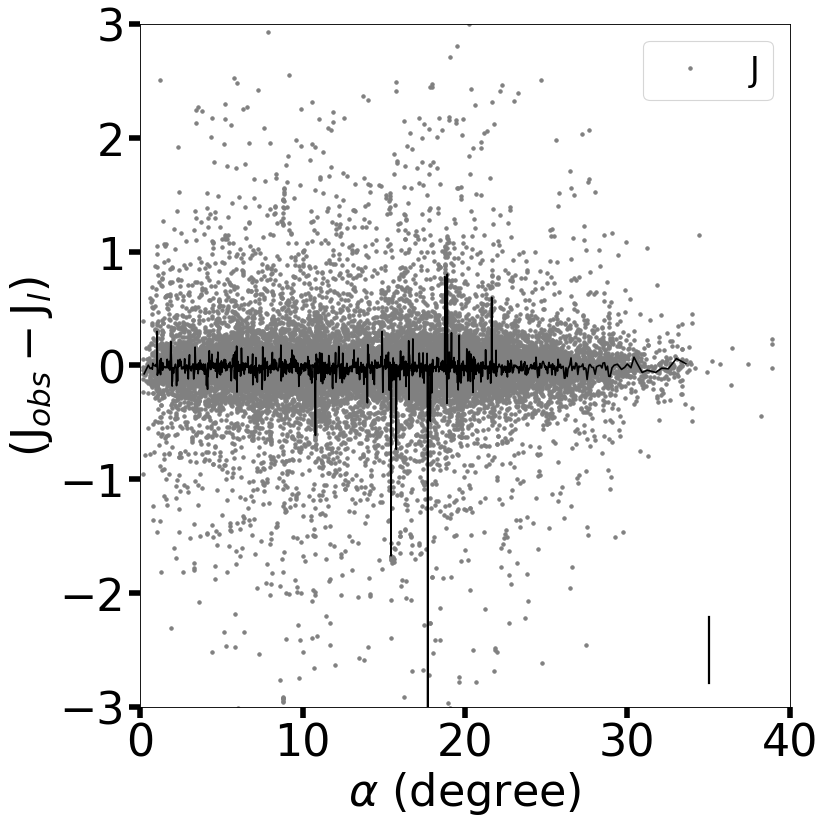}
\includegraphics[width=4.4cm]{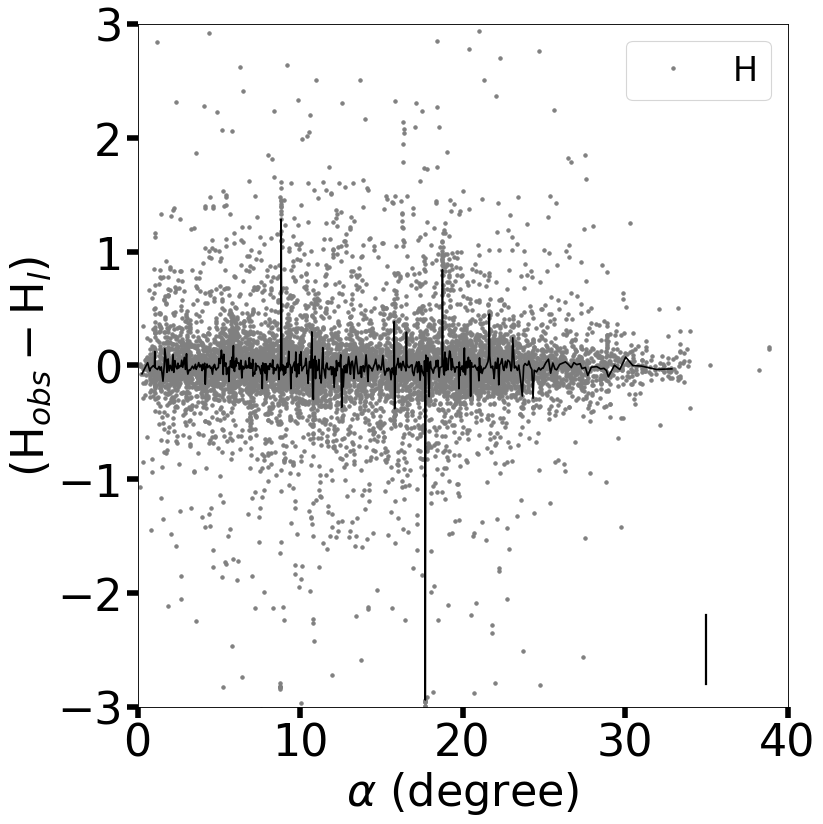}
\includegraphics[width=4.4cm]{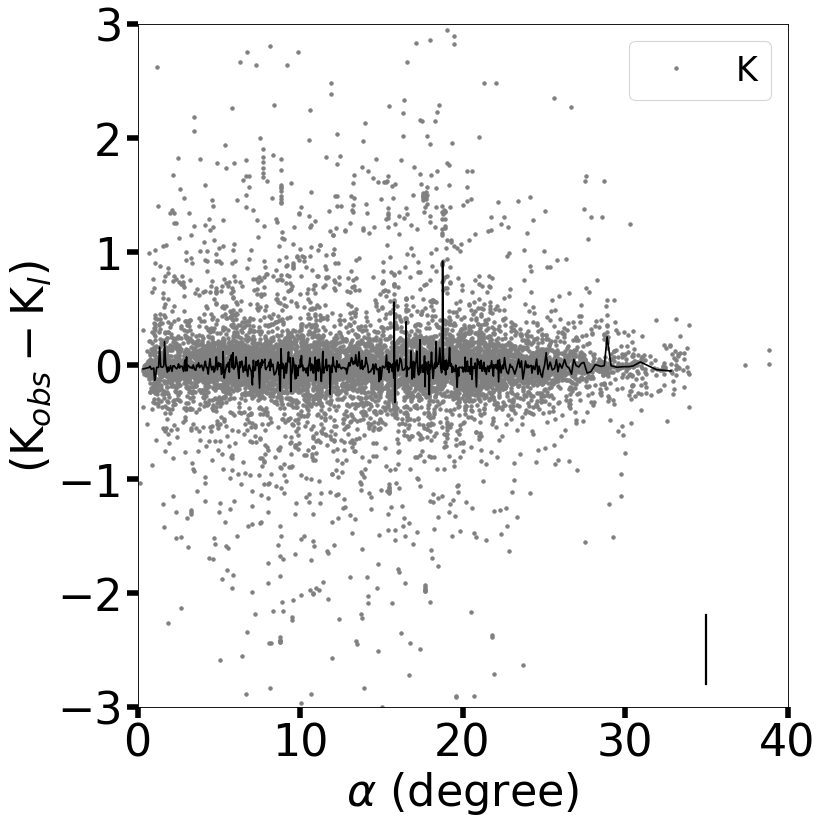}
\caption{Comparison between measured and modeled magnitudes using the linear model. Different panels show data in different filters. The continuous line indicates a running median of 21 points. Typical error bars in the y-axis are shown in the lower-right corner for reference.}
\label{fig:comp2}
\end{figure}
The first one shows the difference between the observed and the HG$_{12}^*$-modeled magnitudes (see Fig. \ref{fig:comp1}, where the plot is cut at $\alpha=40$ deg to remove one point at approximately 80 deg that skews the figure; the same applies to Fig. \ref{fig:comp2}). 
We computed the root mean square (RMS) of each phase curve, obtaining median values of 0.01, 0.02, 0.03, and 0.03 in the Y, J, H, and K filters, respectively. The RMS in the second test, observed minus linear-modeled magnitudes, is slightly lower (0.004, 0.01, 0.01, and 0.01, respectively) (Figs. \ref{fig:comp2}), which may point to an overall better description of the observed data by linear phase curves.

Figures \ref{fig:comp1} and \ref{fig:comp2} display a black line in each panel, representing a smoothed version of the data (computed as the median value of each 21-point interval) to highlight any potential relationship between the differences in magnitudes and the phase angle. Note that in the cases of Fig. \ref{fig:comp1}, there seems to be a residual signal in the smoothed line, particularly in the Y filter. These signals do not appear to be present in the panels of Fig. \ref{fig:comp2}, which supports the idea that in the NIR filters, the photometric phase curves behave more linearly with less prominence of the opposition effect. 

Therefore, we are inclined to accept hypothesis (ii) mentioned above: that the linear model produces fainter magnitudes because the HG$_{12}^*$ model applied to NIR data seems to overestimate the OE region, producing brighter absolute magnitudes. Among the reasons behind these results is that we have poor coverage of the OE region, as only about 12\% of the input observations were obtained with $\alpha < 5$ deg. While this may seem like a substantial amount, it is essential to note that not all of these observations were utilized. Filtering by objects with valid solutions, we find that about 10\% of the objects have at least two data points below 5 deg, while less than 3\% have at least three data points with $\alpha < 5$ deg; thus, the coverage of the OE region tends to be scarce as well. A dedicated program with a few select objects providing dense coverage will be instrumental in this regard.

\begin{figure}[ht]
\centering
\includegraphics[width=4.4cm]{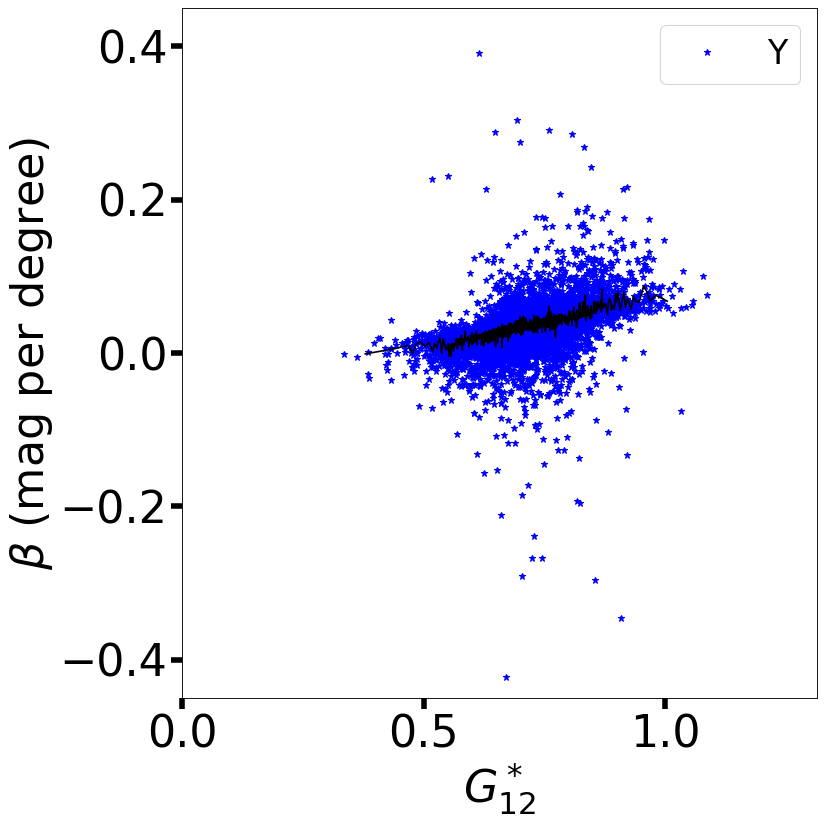}
\includegraphics[width=4.4cm]{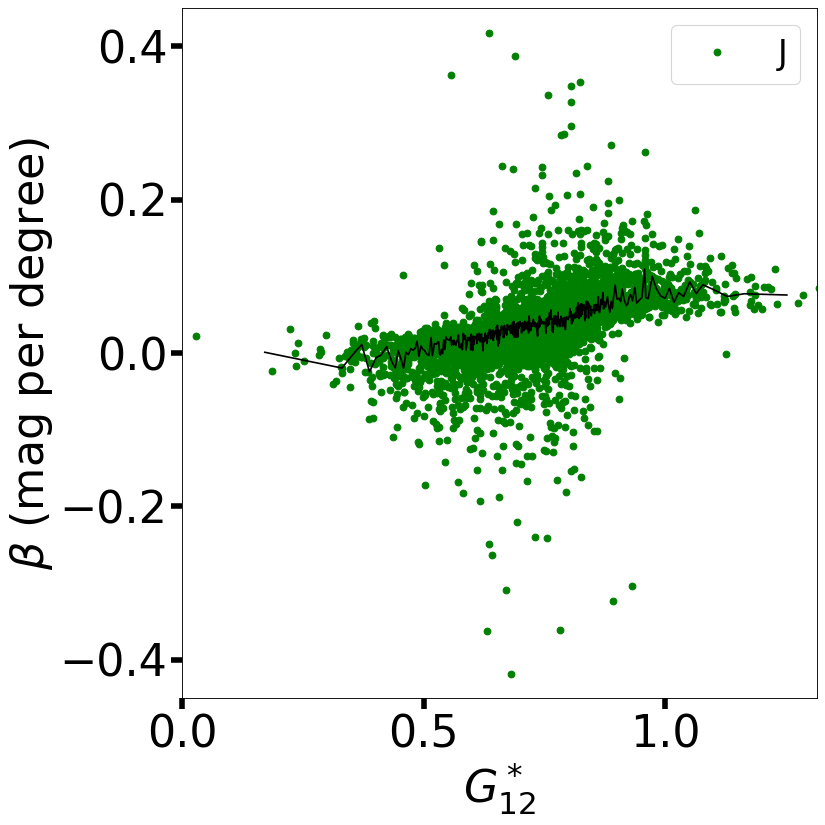}
\includegraphics[width=4.4cm]{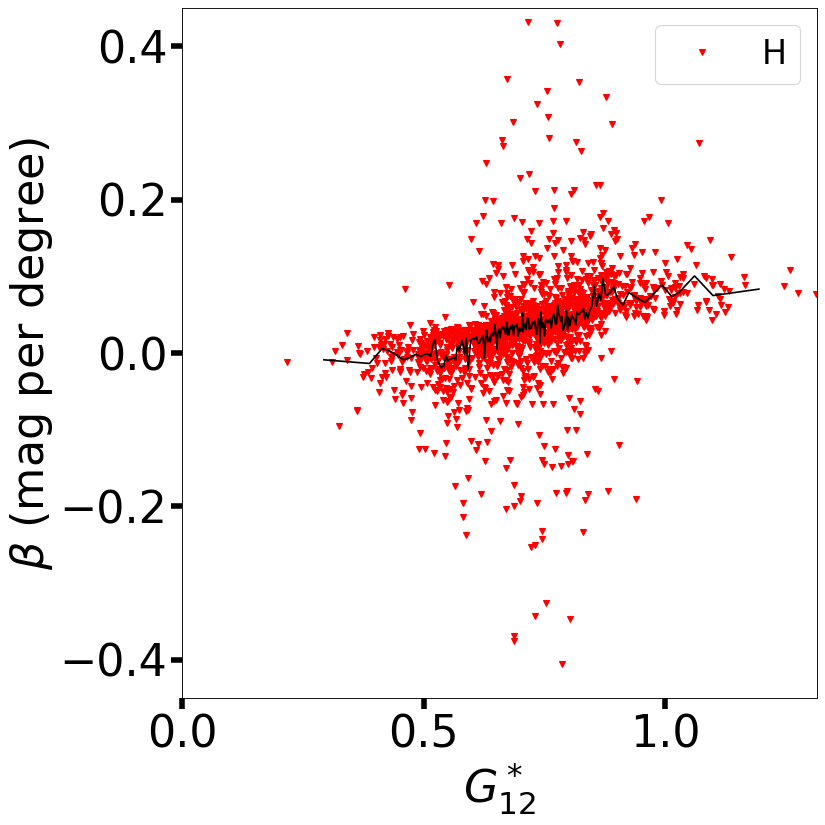}
\includegraphics[width=4.4cm]{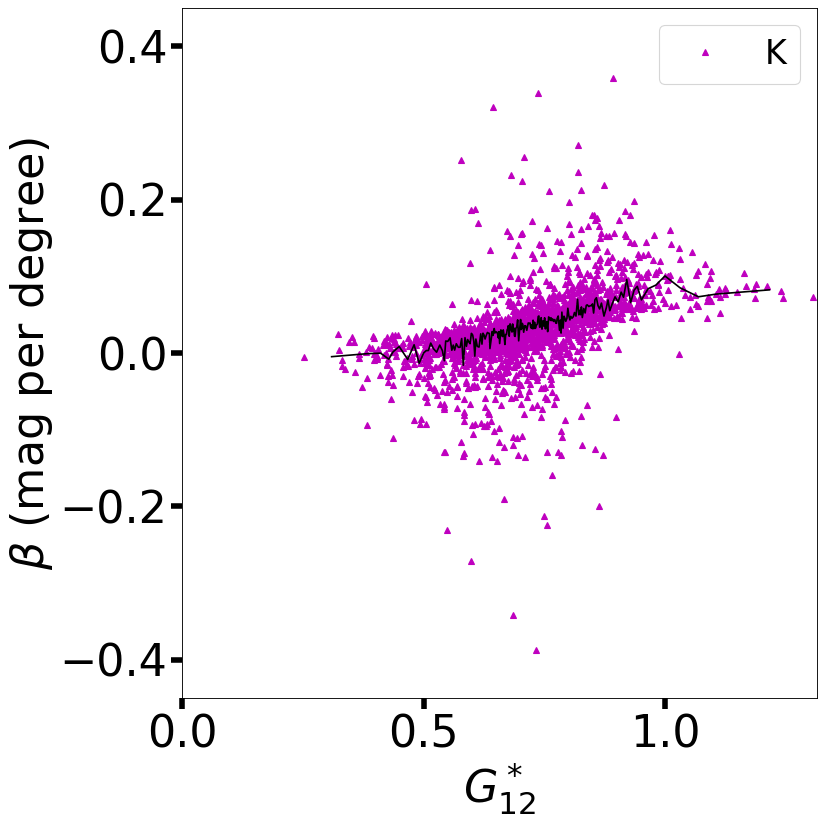}
\caption{Scatter plots showing $G_{12}^*$ vs. $\beta$. $Y$ is shown as the blue line, $J$ is shown in green, $H$ is shown in red, and $K$ is shown in purple. The black continuous line represents a running median window of 11 points, highlighting the relationship between the two sets. Error bars are not shown for clarity. The median error in $G\approx0.5$, while in $\beta\approx0.04$ mag per deg.}
\label{fig:coefs}
\end{figure}

We also checked the relation between the phase coefficients. In Fig. \ref{fig:coefs}, we show $G_{12}^*$ vs. $\beta$ in all filters. In the plots, the data are shown as colored symbols. In contrast, the black continuous line shows a running median window taken every 11 points to highlight the apparent correlation between both samples. The Spearman coefficient is greater than 0.5 in all cases, and the $p_{value}\approx0$ strongly indicates the presence of correlations. In all cases, the behavior is alike: $\beta$ is about 0 deg per mag until $G_{12}^*$ is about 0.5, then it increases until $\beta\approx0.08$ mag per deg at $G_{12}^*\approx0.9$, and then remains almost constant for the remaining range. Note, however, that there is considerably less density of points in the regions of almost constant behavior of $\beta$. 
Figs. \ref{fig:g12-beta} and \ref{fig:coefs} indicate that it may be possible to estimate the values of phase coefficients from the HG$_{12}^*$ model (more time-consuming) using the linear approximation (faster computation). 

Our last test was designed to verify the phase curves for a restricted range of phase angles, specifically targeting observations that lack data at low phase angles, to assess the impact on data fitting. We ran our code with the same setup as above, removing all data below $\alpha=9.5$ deg. We expect that this selection of minimum phase angle should not significantly impact our estimates, as we are dealing in the linear regime of the phase curves \citep{belskayashev2000}. After running the code, we obtained 4\,749 objects with at least one absolute magnitude and 114 with all four values. The complete breakdown of the numbers is shown in Table \ref{tab:objectsHG}. We will call the absolute magnitude and phase coefficients $H_e$ and $\beta_e$, respectively.

The distribution of $H_e$ is similar, though low in absolute numbers, to the one shown in Fig. \ref{fig:disthg12}, see Fig. \ref{fig:distbetas}, bottom right panel. The median values of $H_{\lambda}$ are 15.65 (Y), 13.64 (J), 12.59 (H), and 12.56 (K). In Fig. \ref{fig:distbetas}, right panel, we show the $\beta_e$ distributions; they look similar to the full range of $\alpha$. The median values are slightly different than those in the full-range case (0.031, 0.032, 0.027, and 0.029) mag per deg for (Y, J, H, and K), respectively. The Kruskal-Wallis test does not reject the null hypothesis; therefore, we assume that all distributions are similar. The outliers are 5, 12, 18, and 15, respectively ($|\beta_e|>1.5$ mag per deg).

\begin{figure}[ht]
\centering
\includegraphics[width=4.4cm]{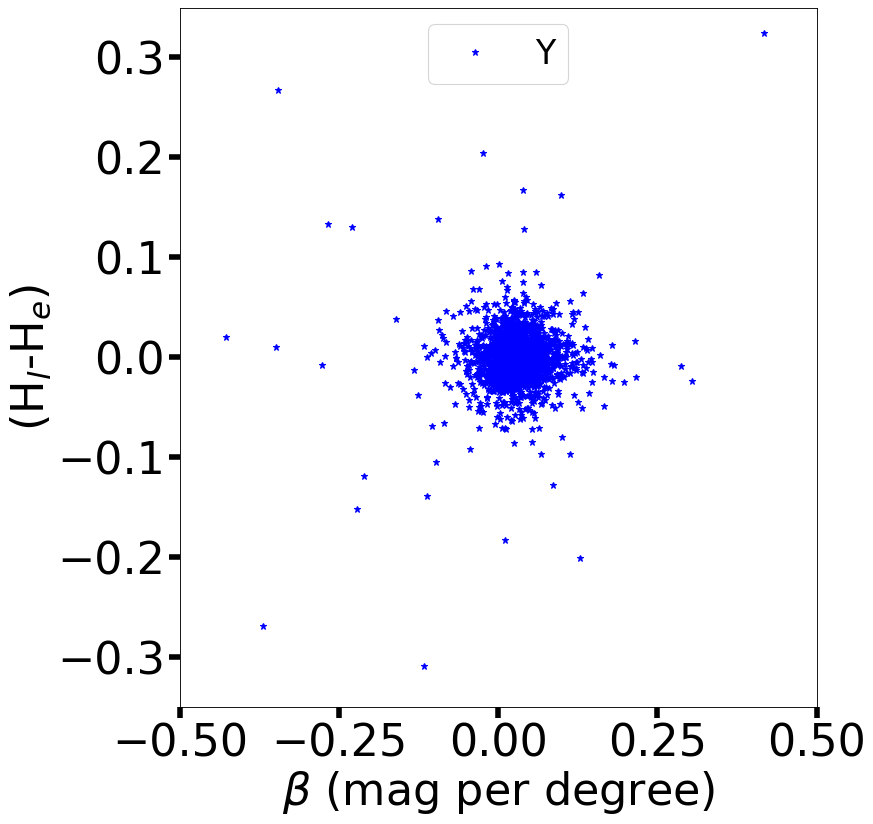}
\includegraphics[width=4.4cm]{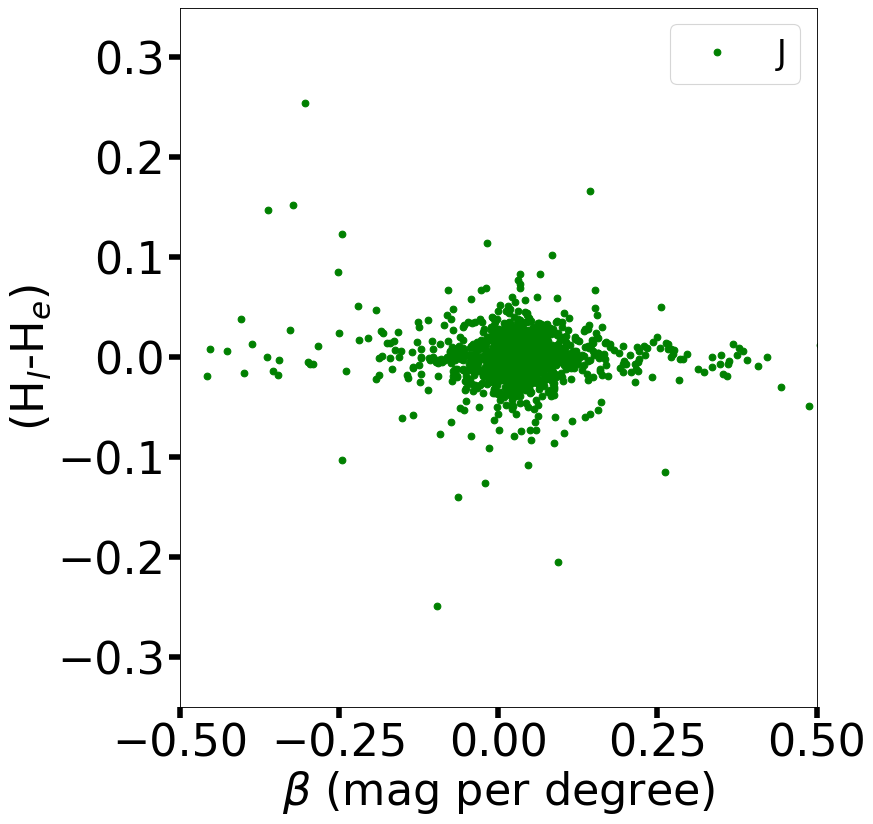}
\includegraphics[width=4.4cm]{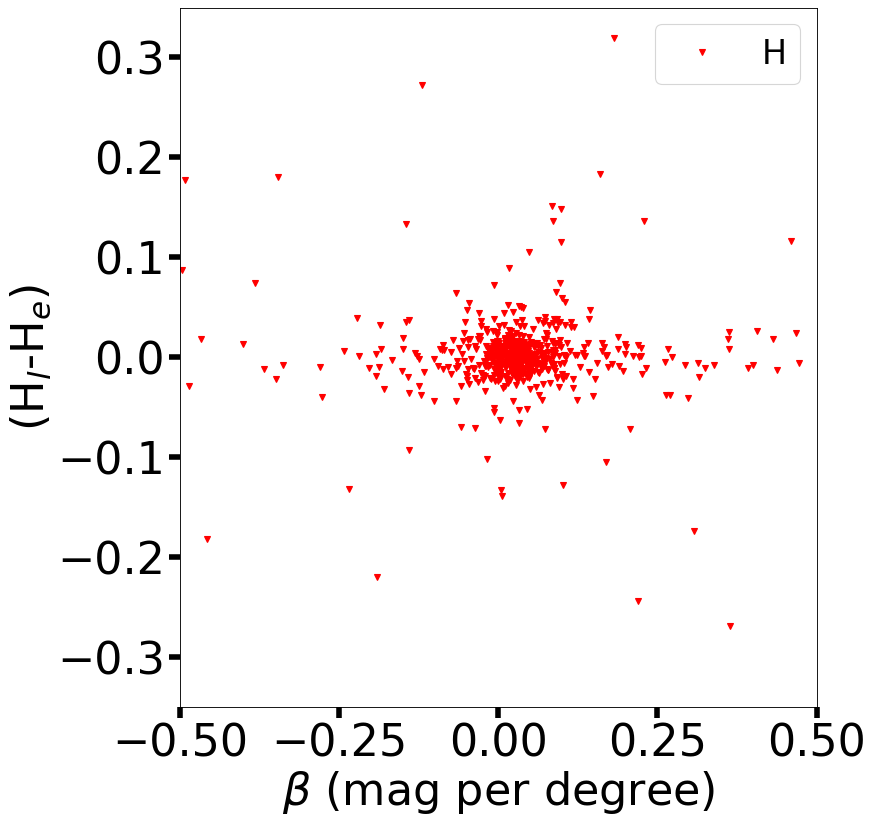}
\includegraphics[width=4.4cm]{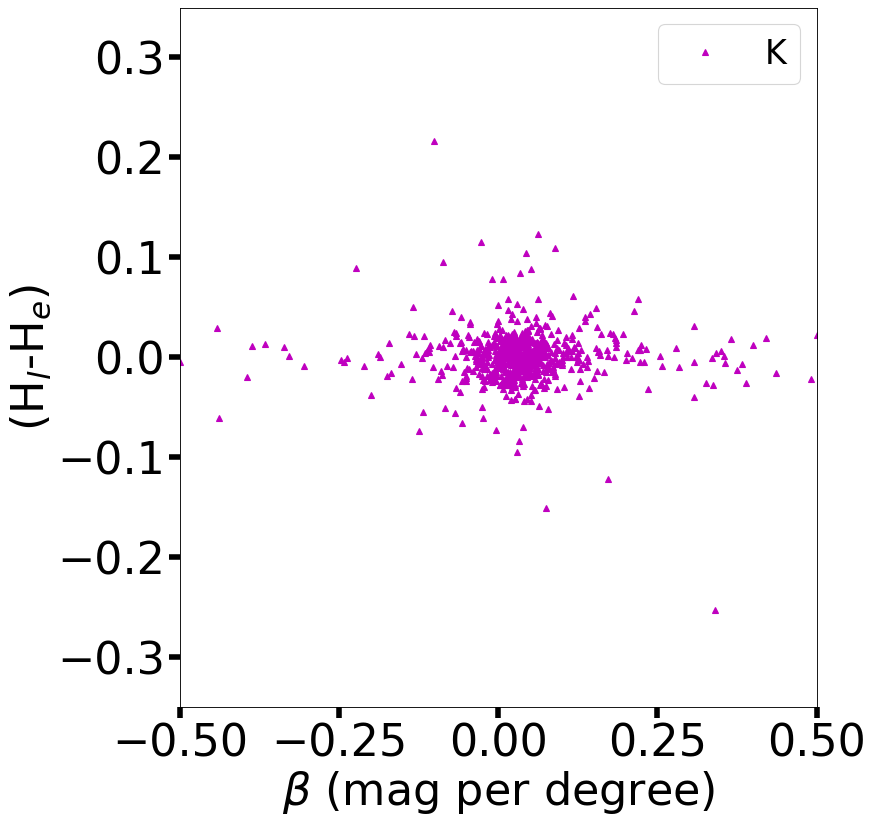}
\caption{Comparison between $H_l$ and $H_e$ as a function of $\beta$. Different panels/colors indicate different filters: $Y$ is shown as the blue line, $J$ in green, $H$ in red, and $K$ in purple. The error bars are not shown for clarity. The typical error on the x-axis is 0.06 mag per deg, while it is about 1.3 mag on the y-axis.}
\label{fig:hle}
\end{figure}

\begin{figure}[ht]
\centering
\includegraphics[width=4.4cm]{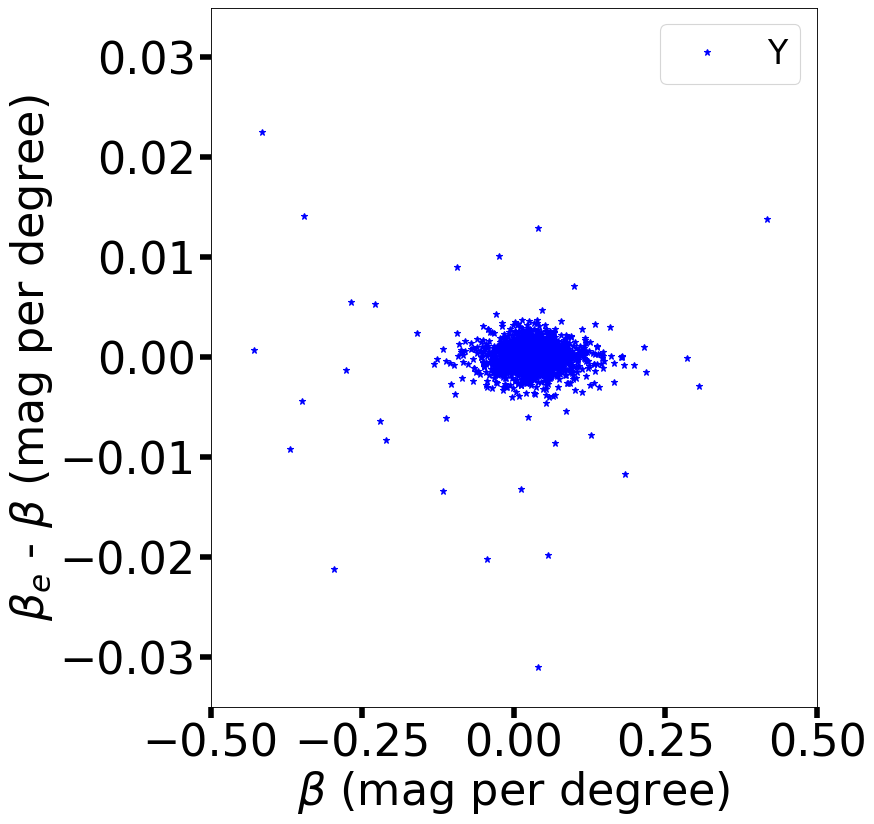}
\includegraphics[width=4.4cm]{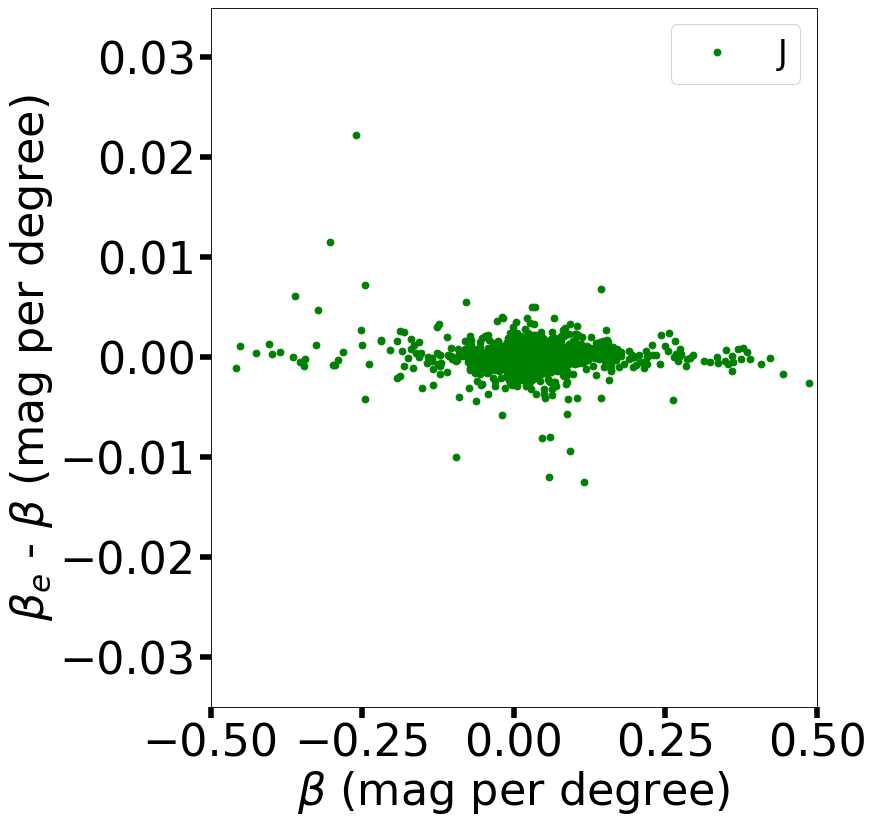}
\includegraphics[width=4.4cm]{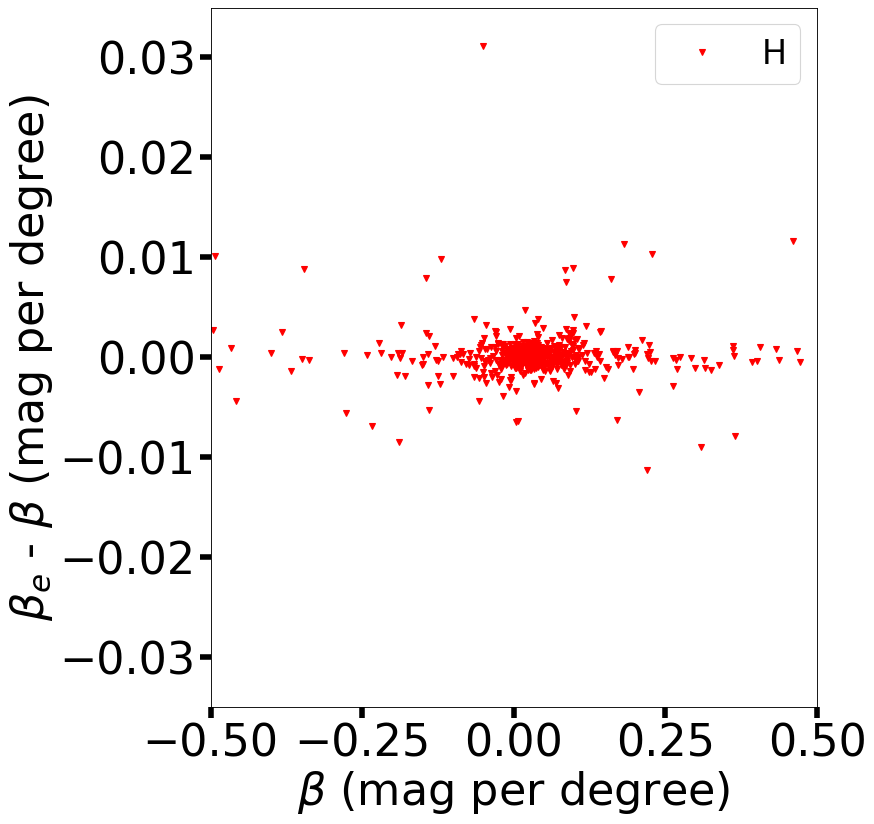}
\includegraphics[width=4.4cm]{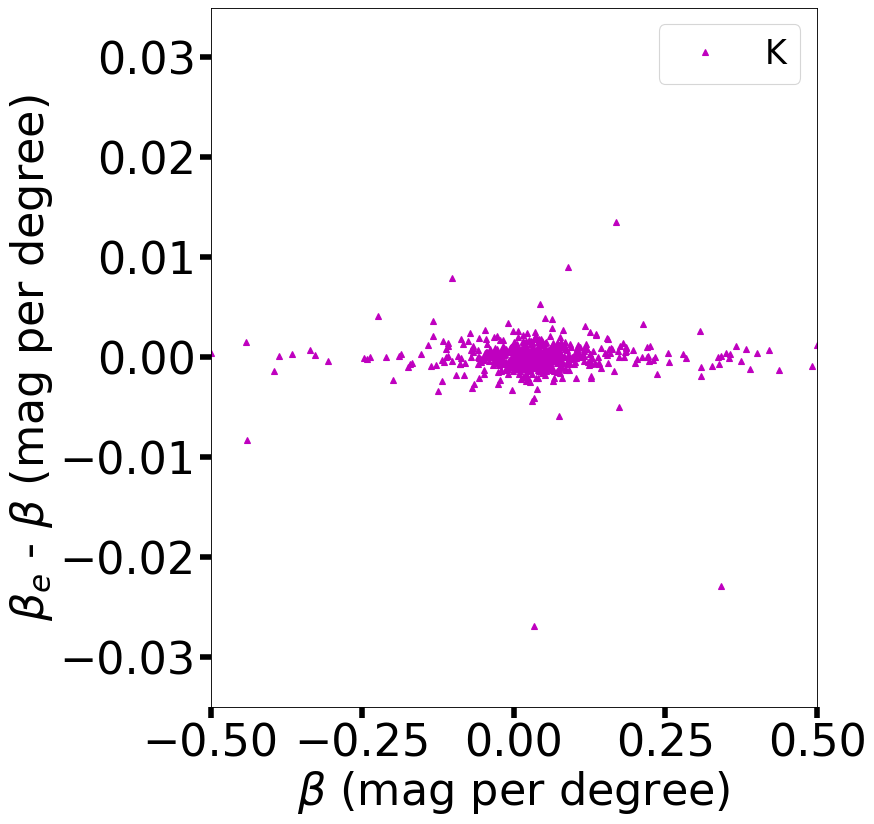}
\caption{Comparison between $\beta$ and $\beta_e$ as a function of $\beta$. Different panels/colors indicate different filters: $Y$ is shown as the blue line, $J$ in green, $H$ in red, and $K$ in purple. The error bars are not shown for clarity. The typical error in the x-axis is 0.06 mag per deg, while it is about 0.08 mag per deg in the y-axis.}
\label{fig:betale}
\end{figure}

We confirm that the difference of using the complete $\alpha$ range or the $\alpha>9.5$ deg is minimal: the median difference is of the order of $10^{-4}$ \textperthousand, or less, noting that $>60\%$ of objects have at least a datum below 9.5 deg. In all filters, 90 \% of the results are within $|(H_{l} - H_{e})| < 3.0 $ mag (see Fig. \ref{fig:hle}), except in the H filter goes up to $|(H_{l} - H_{e})| < 4.6 $ mag. Note that the large uncertainties in the y-axis (median value of 1.3 mag) reflect the wide $P(H,\beta)$ we obtain with our method rather than an estimate of precision. The large difference between using both models come from objects with low number of observations, where removing one datum below 9.5 deg produces a large variation in the slope of the phase curve and therefore the estimated absolute magnitude.
Therefore, in principle, having a restricted range of $\alpha$ should not strongly impact estimating the absolute magnitude because the linear model is well adapted for this. The same is indicated for the phase coefficients, as displayed in Fig. \ref{fig:betale}.
We conclude that using a linear model to fit data with $\alpha > 9.5$ should not differ significantly from doing so with a complete range of phase angles for NIR data. 

\subsection{Phase coloring}

Recent results point out that the usually nicknamed {\it phase reddening} effect is a coloring effect that may go either way in affecting the colors \cite[and therefore the reflectances][]{ayala2018,alcan2024AA,wilawer2024MNRAS,colazo2025Icarus}. However, these works dealt with phase curves in the visible spectrum. We extend the analysis into the near infrared with our new results. In this part of the work whenever reading $H_{\lambda}$ it should be understood as $H_{l_{\lambda}}$, unless explicitly mentioned otherwise. We obtained remarkably similar results to those reported in \cite{ayala2018} for trans-Neptunian Objects and \cite{alcan2022b} for asteroids. In these cases, the authors restricted the phase angle range to that accessible by TNOs and Centaurs, whereas in our case, we use the linear model with no limit on phase angle coverage.
\begin{figure}[ht]
\centering
\includegraphics[width=4.4cm]{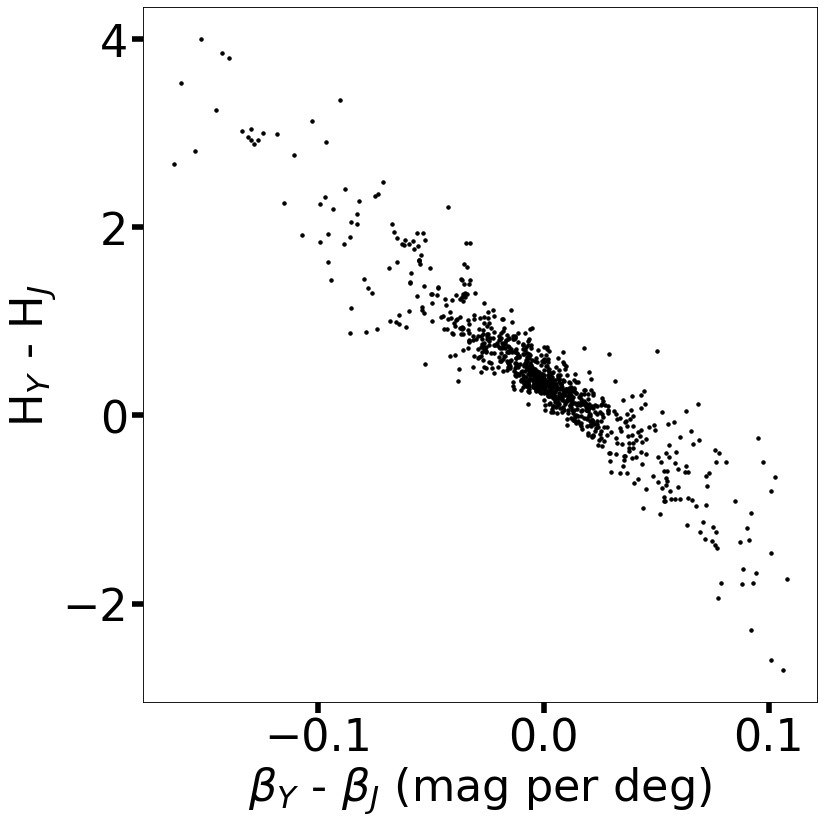}
\includegraphics[width=4.4cm]{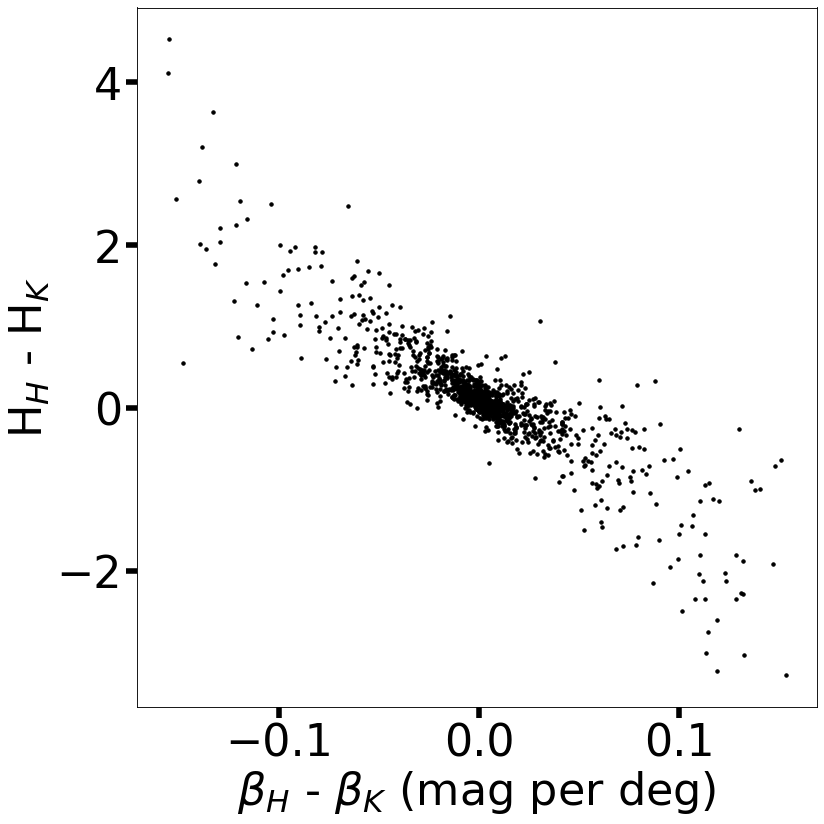}
\caption{Colors versus relative phase coefficients. The left panel shows $Y-J$, while the second panel shows $H-K$. Error bars are not shown for clarity.}
\label{fig:HsBs}
\end{figure}

The figures above indicate that redder objects at opposition tend to become bluer with increasing phase angle, while neutral-to-blue objects tend to become redder with increasing phase angle. As discussed in the mentioned works, we do not yet have a definite answer as to why this happens, but we are making strides in understanding that porosity, packing, and wavelength dependence of the albedo may be behind it. These results are valid within the range of $\alpha$ where the models were fit, and, accepting the limitations of a simple linear model, there is evidence of a general phase coloring effect, rather than just phase reddening.

\subsection{Uncertainties}

To understand the uncertainty budget of our results, it is necessary to grasp where they come from: (i) We combined four different catalogs with different depths, photometric filters (Fig. \ref{fig:filters}), and photometric systems (AB vs. Vega). Regarding the depth, there is not much we can do; the objects whose data could be combined have already been combined. The transformations between filters and systems were described in Sect. \ref{sec:data} and the estimated offsets were given in the same section. The most affected data is in the Y filter, with a difference of roughly 0.6 mag between MOVIS and DES, while the JHK data conversions have an impact of about 0.05 mag. (ii) Data are snapshots at different epochs, and we do not know the rotational phase at which they were taken. To account for these unknown rotational phases, we applied a Monte Carlo sorting technique from a probability distribution $P(\Delta m)$ for all asteroids with a wide FWHM (usually about 0.5 mag, see Fig. 2 in \citealt{alcan2024AA}) and a total range of about three mag. (iii) Finally, there are some effects we are not including in our modeling, such as the changing aspect angle and different apparitions (as in \citealt{robinson2024-atlaspc} or \citealt{carry2024}). Nevertheless, our approach utilizes the light-curve amplitudes from \cite{wagner2019}, which include data from different apparitions and phase angles. Thus, our probability distributions $P_A(\Delta m)$ average out this effect on a large basis.

\subsection{Can these results be used with Euclid data?}

The Euclid mission will map about 15\,000 sq. deg of the sky, using a particular set of YJH filters, observing in almost quadrature with the Sun (solar elongation about 90 deg). This implies that the serendipitous observations of small bodies will occur at relatively large phase angles in the Main Belt ($\alpha\approx18.5$ deg at about 3 AU \citealt{carry2018euclid,euclid2025_paper1}). Therefore, estimating absolute magnitudes with a restricted phase-angle coverage is crucial. In Sect. \ref{sec:largea} we showed that the linear model works well even if no data is present below 9.5 deg, but in the Euclid data case, we are far from that region; therefore, our results
may not hold due to the limited phase angle coverage. On the other hand, due to the observational strategy, it will be rare for the same object to be observed more than once. Additionally, Euclid's photometric system differs from those described in this section.

Nevertheless, using our results to estimate absolute magnitudes from a single Euclid observation may still be possible. Let us assume an observation with $J_{euclid}=15.0\pm0.1$ at $\alpha=18.5$ deg. From the results shown above, we can create the prior $p(H_l,\beta)$ using J-filter data and the linear model. For this exercise, we will assume that the MOVIS photometric system (our chosen reference) is identical to Euclid's. We compute the posterior probability
\begin{equation}
    p(H_l, \beta | J_{euclid}) \propto p(J_{euclid} | H_l, \beta) p(H_l, \beta),
\end{equation}
where $p(J_{euclid}|H_l,\beta)$ is the probability of measuring $J_{euclid}$ from our model. We will use a Gaussian likelihood to estimate our posterior probability and disregard the normalization factor, which is irrelevant in this case.

\begin{figure}[ht]
\centering
\includegraphics[width=4.4cm]{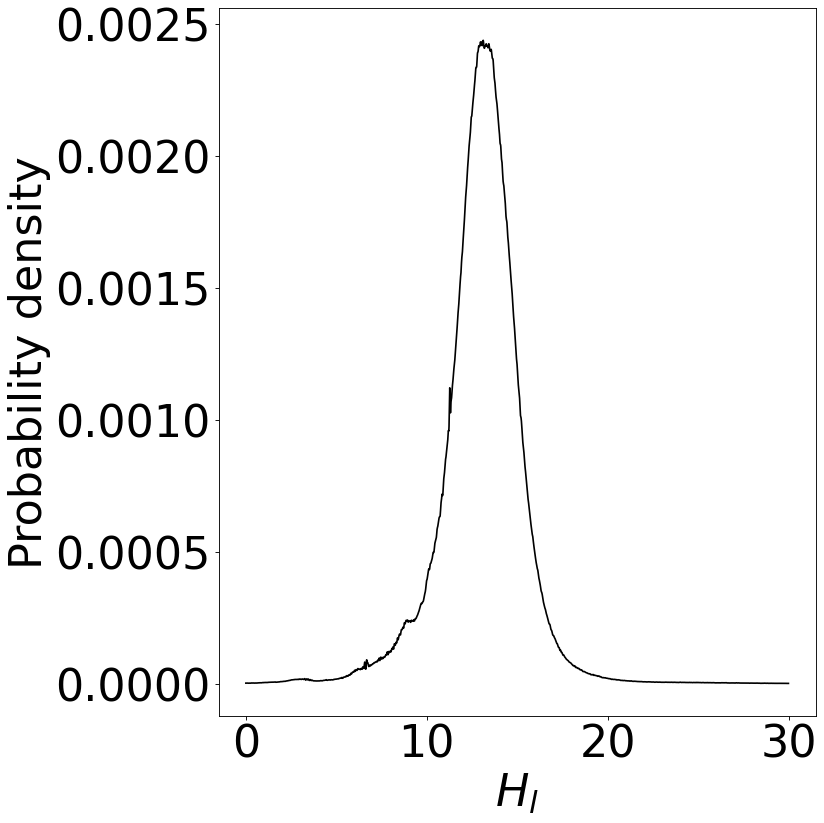}
\includegraphics[width=4.4cm]{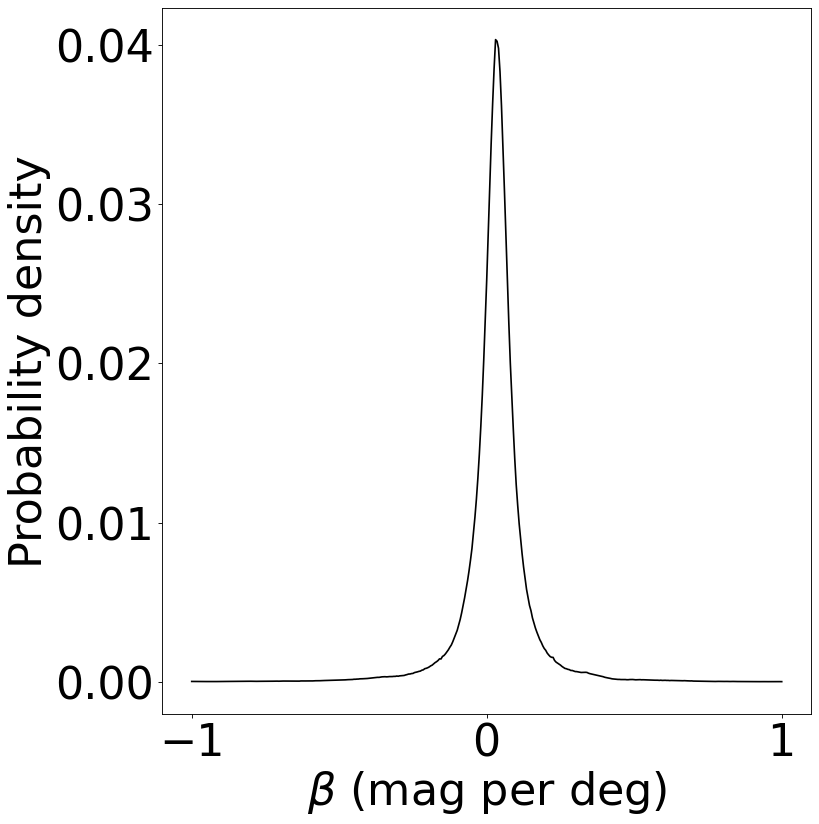}
\caption{Marginal distributions of $p(H_l,\beta|J_{euclid})$. Left: Absolute magnitudes; right: Phase coefficients.}
\label{fig:marginals}
\end{figure}

The resulting values of $H_{euclid}$ and $\beta_{euclid}$ are estimated as the median of the marginal distributions of $p(H_l,\beta|J_{euclid})$, and the uncertainty margins as the 16th and 84th percentiles (see Fig. \ref{fig:marginals}). We obtained $H_{euclid}=13.2^{+1.6}_{-1.8}$ and $\beta_{euclid}=0.040_{-0.065}^{+0.055}$ mag per deg. These values, despite the large uncertainties, show that it is possible to estimate absolute magnitudes from a single Euclid Consortium observation at large $\alpha$ when a suitable prior is available.

\section{Conclusions}

In this work, we computed absolute magnitudes in near-infrared filters, Y, J, H, and K, for $>10\,000$ small bodies. The methodology used was established elsewhere; therefore, our absolute magnitudes and phase coefficients accurately represent the phase curves of these objects. Our results indicate that a linear function can accurately describe the phase curves and that it is unnecessary to use complex, non-linear models, which may overestimate the OE, resulting in brighter absolute magnitudes than the actual ones (see also Appendix \ref{appA}). Because a linear function well represents the curves, the impact of using a limited range in phase angle (for instance, using $\alpha>9.5$ deg) is minimal.

On the other hand, we also confirm the effect of phase coloring happening in the near-infrared. The same behavior is observed in the visible for asteroids and trans-Neptunian objects, which is detected in near-infrared photometry: redder (bluer) objects at opposition tend to become bluer (redder) with increasing phase angles. Indeed, the relations shown in Fig. \ref{fig:HsBs} reflect the linear model used; the shape would be different if other photometric models were employed, such as those in \cite{alcan2024AA}. Unfortunately, the number of objects with NIR magnitudes is comparatively small when seen against photometric surveys in the visible. Note that we needed to join four catalogs to reach critical mass (results for $> 10\,000$ objects) and still remain an order of magnitude (at least) below any survey in the visible. 

The results presented in this work confirm that the data from the Euclid Consortium can be transformed into absolute magnitudes using a simple linear mode and our results as prior information, and provided a conversion between the photometric systems is known.

\paragraph{Take away messages:}
\begin{itemize}
    \item A linear model is sufficient to fit NIR magnitudes.
    \item Euclid data can provide useful estimates of absolute magnitudes, provided suitable priors exist.
    \item Phase coloring also appears in phase curves in the NIR.
\end{itemize}

%\begin{acknowledgments}
\section*{Acknowledgments}
\smallskip
\noindent
AAC acknowledges financial support from the Severo Ochoa grant CEX2021- 001131-S funded by MCIN/AEI/10.13039/501100011033. AAC, JLR, MC, RD, and DM acknowledge financial support from the Spanish project PID2023-153123NB-I00, funded by MCIN/AEI. MC was supported by grant No. 2022/45/B/ST9/00267 from the National Science Centre, Poland. All figures in this work can be reproduced using the Google Colab Notebook \url{https://colab.research.google.com/drive/1NIjqIpwAYj8e7t8ggp94CDD\_cb4Yqqy2?usp=sharing}. All data are available at \url{https://osf.io/cfxq2} and at the CDS.\\
This publication makes use of data products from the Two Micron All Sky Survey, which is a joint project of the University of Massachusetts and the Infrared Processing and Analysis Center/California Institute of Technology, funded by the National Aeronautics and Space Administration and the National Science Foundation.\\
This project used public archival data from the Dark Energy Survey (DES). Funding for the DES Projects has been provided by the U.S. Department of Energy, the U.S. National Science Foundation, the Ministry of Science and Education of Spain, the Science and Technology FacilitiesCouncil of the United Kingdom, the Higher Education Funding Council for England, the National Center for Supercomputing Applications at the University of Illinois at Urbana-Champaign, the Kavli Institute of Cosmological Physics at the University of Chicago, the Center for Cosmology and Astro-Particle Physics at the Ohio State University, the Mitchell Institute for Fundamental Physics and Astronomy at Texas A\&M University, Financiadora de Estudos e Projetos, Funda{\c c}{\~a}o Carlos Chagas Filho de Amparo {\`a} Pesquisa do Estado do Rio de Janeiro, Conselho Nacional de Desenvolvimento Cient{\'i}fico e Tecnol{\'o}gico and the Minist{\'e}rio da Ci{\^e}ncia, Tecnologia e Inova{\c c}{\~a}o, the Deutsche Forschungsgemeinschaft, and the Collaborating Institutions in the Dark Energy Survey.
The Collaborating Institutions are Argonne National Laboratory, the University of California at Santa Cruz, the University of Cambridge, Centro de Investigaciones Energ{\'e}ticas, Medioambientales y Tecnol{\'o}gicas-Madrid, the University of Chicago, University College London, the DES-Brazil Consortium, the University of Edinburgh, the Eidgen{\"o}ssische Technische Hochschule (ETH) Z{\"u}rich,  Fermi National Accelerator Laboratory, the University of Illinois at Urbana-Champaign, the Institut de Ci{\`e}ncies de l'Espai (IEEC/CSIC), the Institut de F{\'i}sica d'Altes Energies, Lawrence Berkeley National Laboratory, the Ludwig-Maximilians Universit{\"a}t M{\"u}nchen and the associated Excellence Cluster Universe, the University of Michigan, the National Optical Astronomy Observatory, the University of Nottingham, The Ohio State University, the OzDES Membership Consortium, the University of Pennsylvania, the University of Portsmouth, SLAC National Accelerator Laboratory, Stanford University, the University of Sussex, and Texas A\&M University.\\
Based in part on observations at Cerro Tololo Inter-American Observatory, National Optical Astronomy Observatory, which is operated by the Association of Universities for Research in Astronomy (AURA) under a cooperative agreement with the National Science Foundation.\\
This research has made use of the SVO Filter Profile Service "Carlos Rodrigo", funded by MCIN/AEI/10.13039/501100011033/ through grant PID2023-146210NB-I00. This work used \url{https://www.python.org/}, \url{https://numpy.org} \citep{numpy}, \url{https://www.scipy.org/}, Matplotlib \citep{hunte2007}, and \url{https://scikit-posthocs.readthedocs.io}. \\
The text was revised using Grammarly and double-checked by the authors.
%\end{acknowledgments}

%\bibliographystyle{aa} % style aa.bst
\bibliography{paper} % your references Yourfile.bib

\begin{appendix}
\section{Summary of conversions and solar colors}\label{appB}

$\bullet$ 2MASS to MOVIS\\
$J_M = J_2 - 0.077 (J-H)_2,  \\
H_M = H_2 + 0.032 (J-H)_2,  \\
K_M = K_2 + 0.010 (J-K)_2.$ \\
Solar Colors: $(J-H)_2 = 0.286$ and $(J-K)_2 = 0.362$.

\smallskip
\noindent
$\bullet$ UHS to 2MASS\\
$J_2 = J_U + 0.065 (J-H)_U,  \\
H_2 = H_U + 0.030 - 0.070 (J-H), \\
K_2 = K_U - 0.010 (J-K)_U,  $\\
Solar Colors:  $(J-H)_U=0.35$ and $(J-K)_U = 0.40$.

\smallskip
\noindent
$\bullet$ DES to MOVIS\\
$Y_M = Y_D - 0.6$

\section{Direct comparisons}\label{appA}
\begin{figure}[h]
\centering
  \includegraphics[width=4cm]{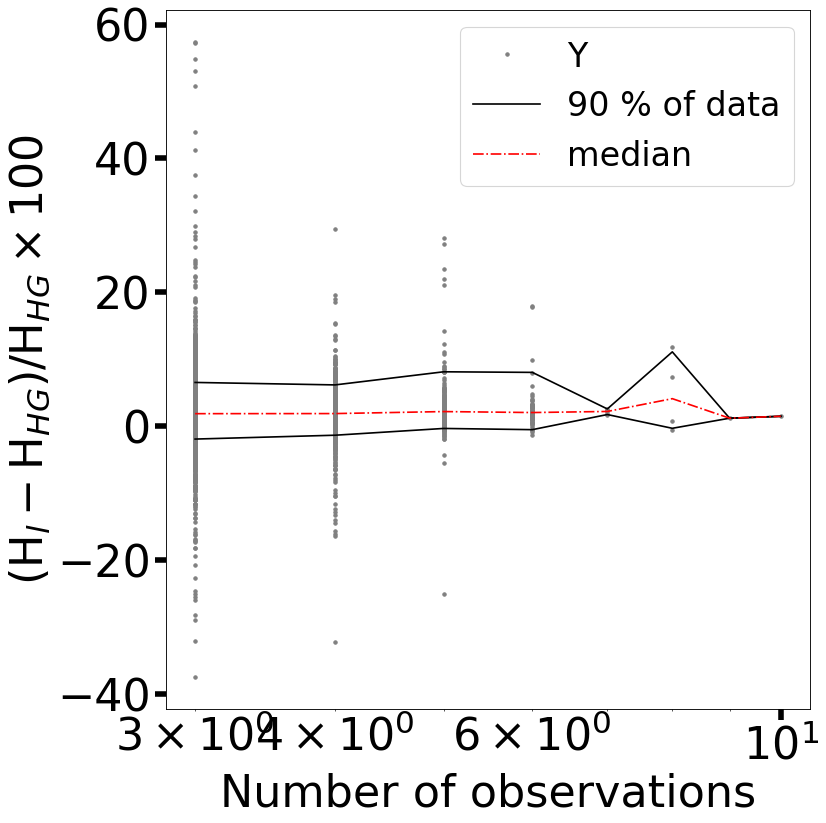}
  \includegraphics[width=4cm]{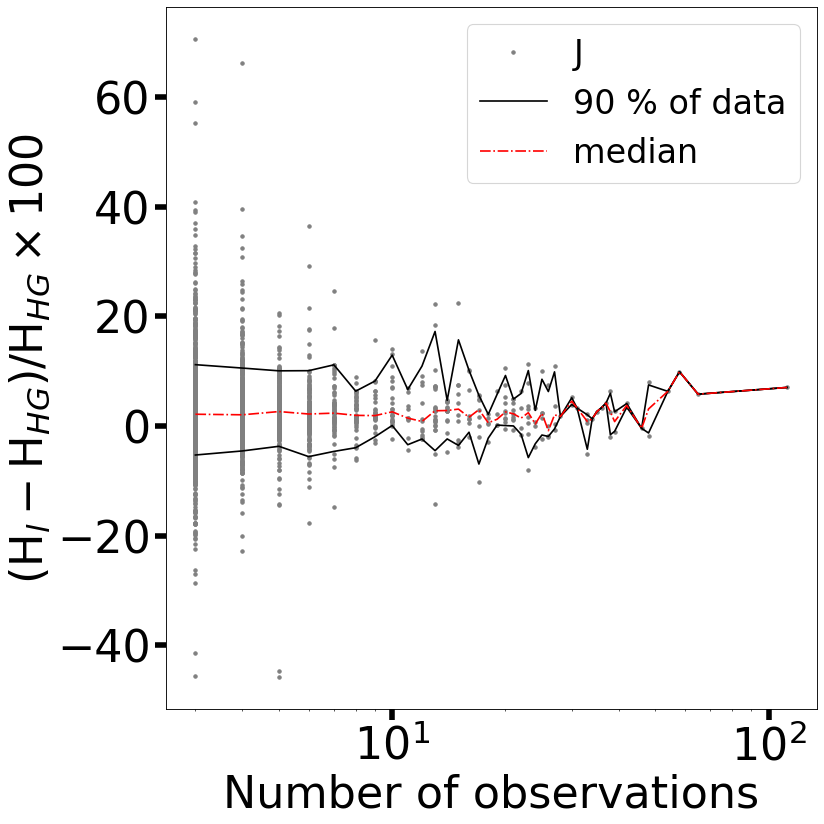}
  \includegraphics[width=4cm]{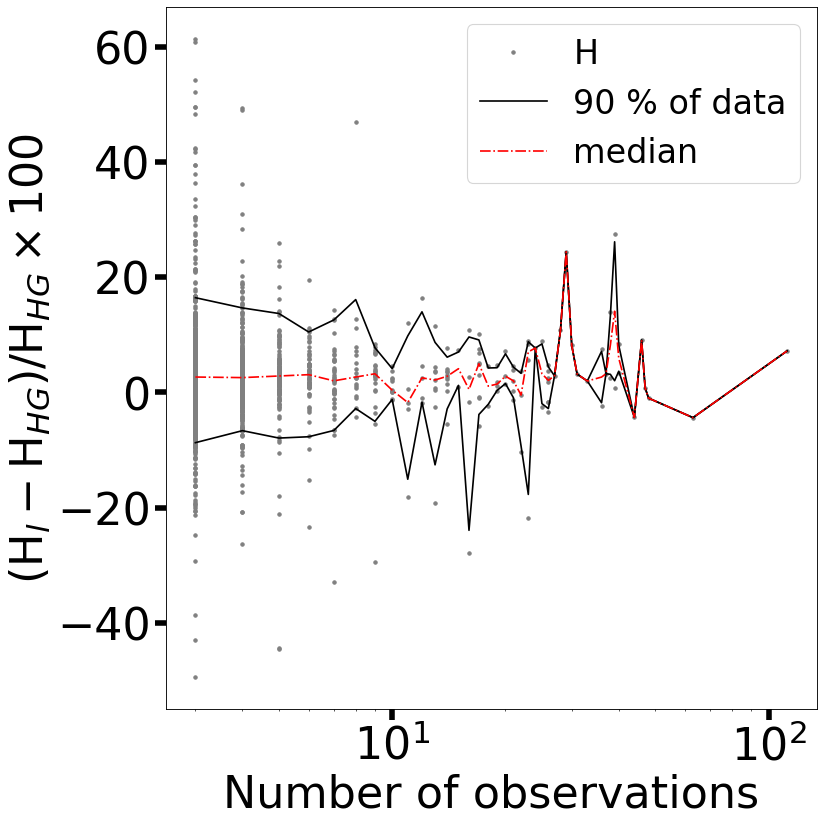}
  \includegraphics[width=4cm]{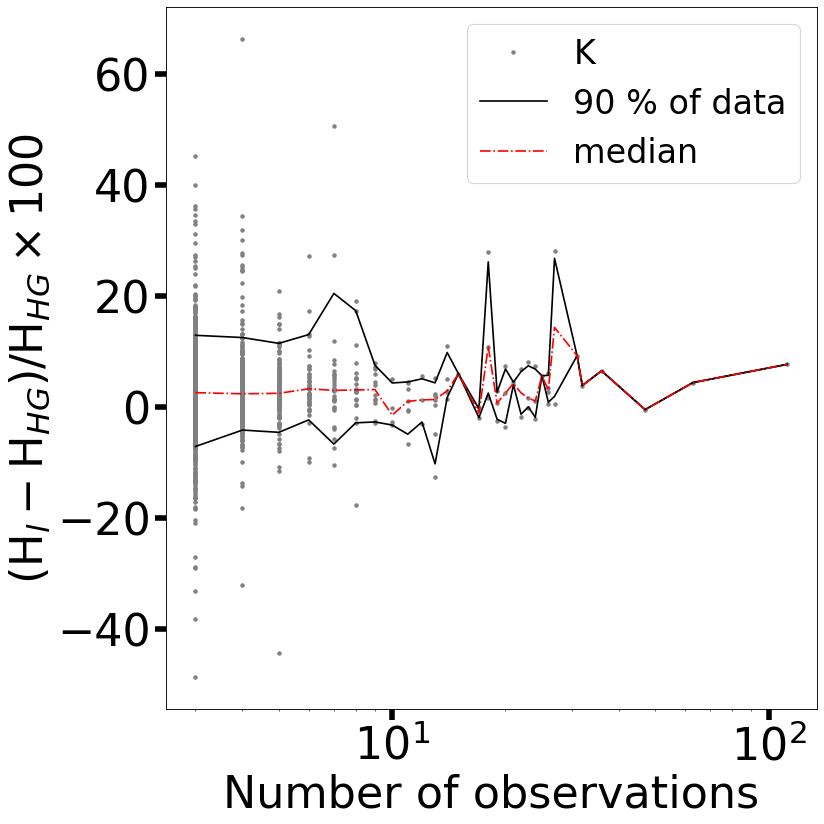}
\caption{$(H_{l}-H_{HG_{12}^*})/H_{HG_{12}^*}\times100$ vs. number of observations. Each panel is labeled with the filter name. The continuous black line indicated the limits of 90\% of the data, while the red dot-dashed line indicated the median curve.}\label{fig:a01}%
\end{figure}

We directly compared absolute magnitudes obtained with the HG$_{12}^*$ and linear systems vs. the number of observations per object (Fig. \ref{fig:a01}). There is a large spread, reaching values as high as about 60 \% of the difference in extreme cases, especially for a low number of observations. Still, we checked that the 90 \% of the data (black curve in the figure) is always less than 25-30 \%, and smaller in the case of the Y filter (maximum spread about 13 \%). From the figures, it is also clear that the HG$_{12}^*$ absolute magnitude overestimates the value with respect to the linear model by 2 to 3 \%, possibly because it tries to fit the OE region.
\smallskip

\begin{figure}[h]
\centering
  \includegraphics[width=4cm]{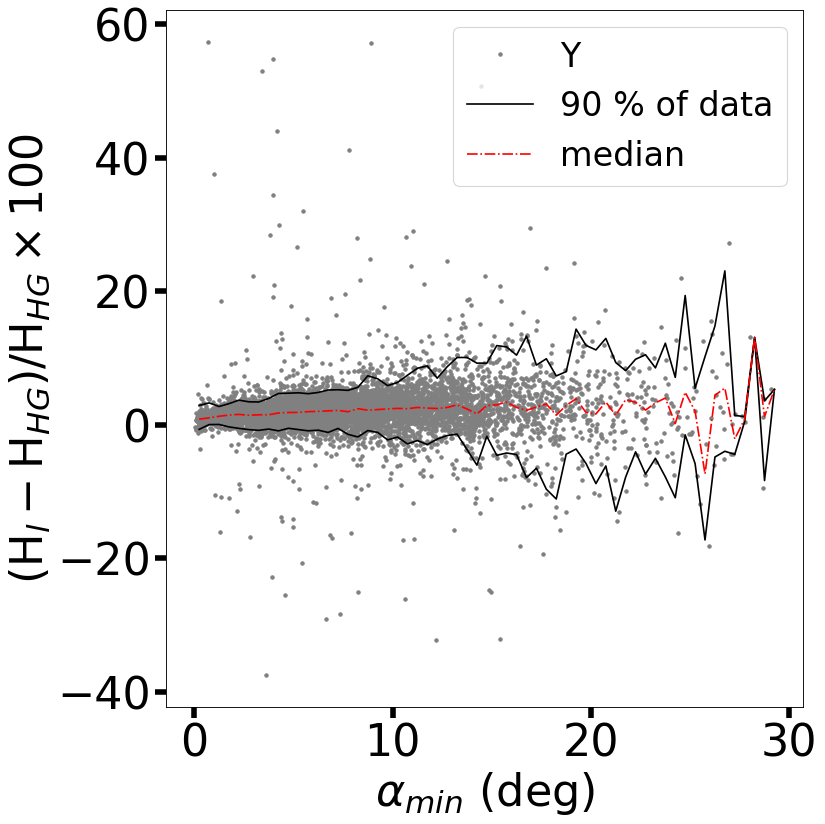}
  \includegraphics[width=4cm]{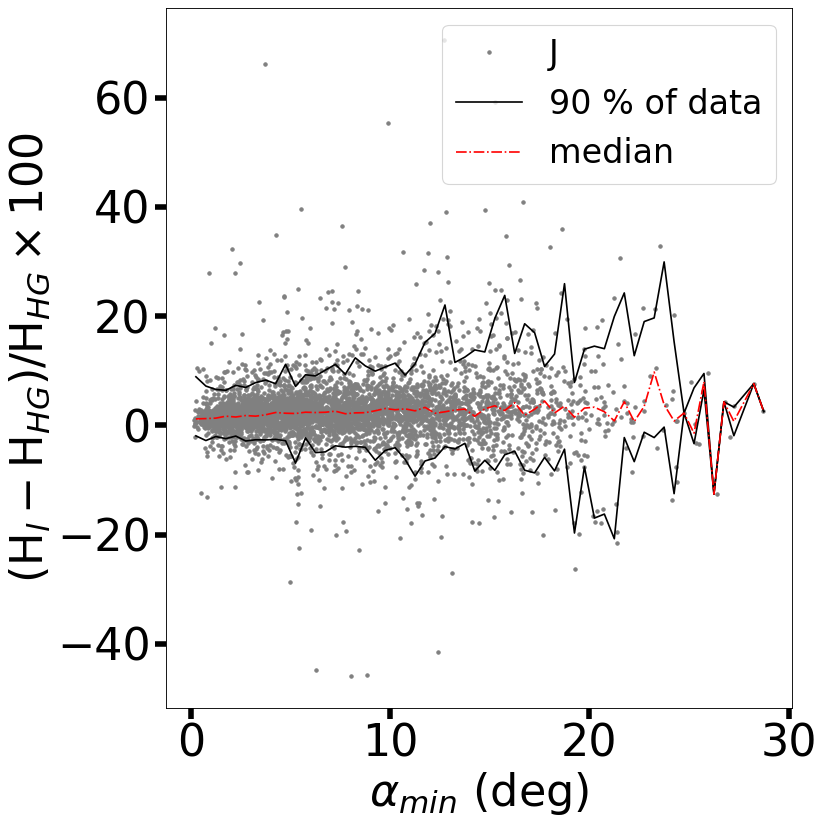}
  \includegraphics[width=4cm]{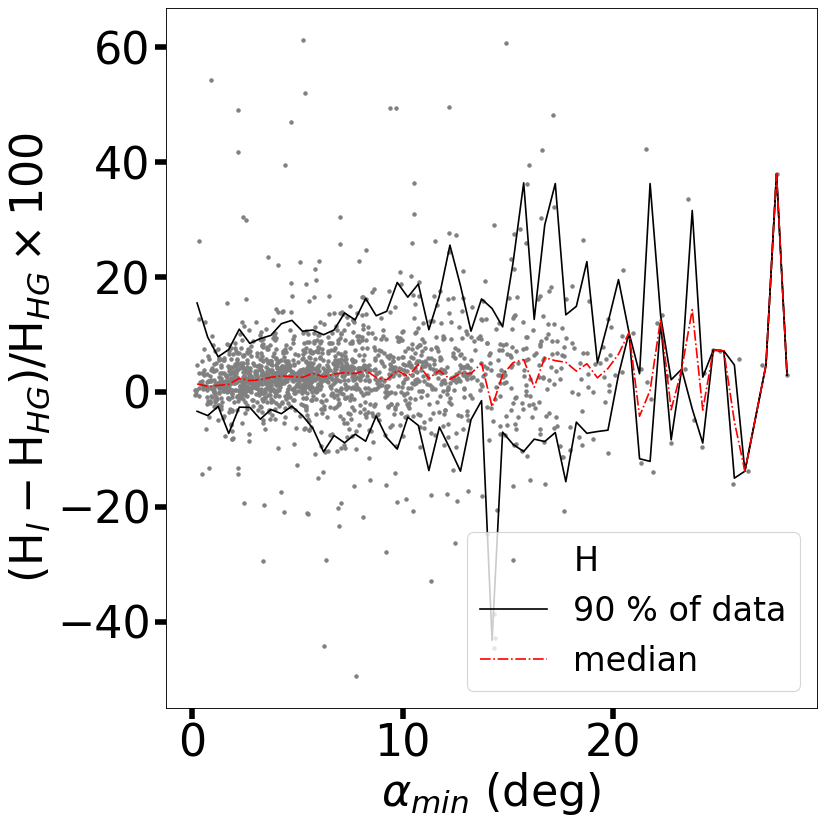}
  \includegraphics[width=4cm]{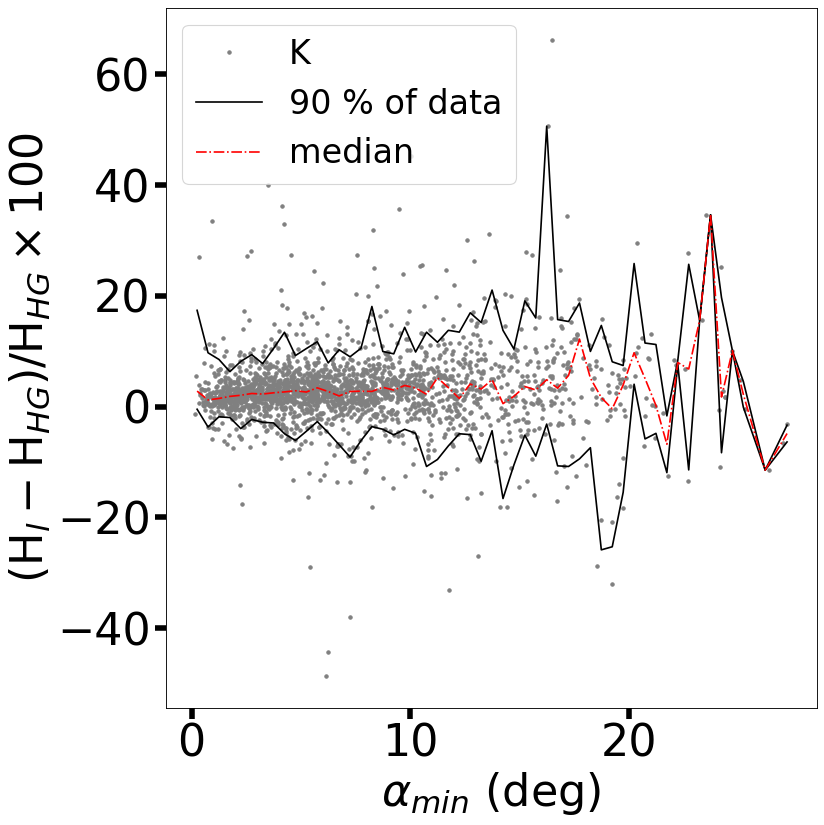}
\caption{$(H_{l}-H_{HG_{12}^*})/H_{HG_{12}^*}\times100$ vs. $\alpha_{min}$. Each panel is labeled with the filter name. The continuous black line indicated the limits of 90\% of the data, while the red dot-dashed line indicated the median curve.}\label{fig:a02}%
\end{figure}

Next, we make the same comparison, but with the $\alpha_{min}$, i.e., the minimum phase angle observed per object. Figure \ref{fig:a02} shows the results. The black lines have a median spread between 15 and 20 \%, depending on the filter used (smaller for Y, larger for H), and the median values of the difference are about 2.5 to 3 \%, as seen above. In this plot, we can see that systematically, the median curve (in red) tends to increase with increasing $\alpha_{min}$, implying that the linear absolute magnitudes are less bright than the HG$_{12}^*$ ones. This behavior may be expected if the HG$_{12}^*$ model is overestimating the OE: Whenever there is data at low-$\alpha$ the linear and the HG$_{12}^*$ model do not depart too much from each other, but in the absence of data below 5 to 7 deg, the HG$_{12}^*$ model may tend to over-fit the OE region, producing brighter absolute magnitudes.

\end{appendix}
\end{document}